%% file: hmd.tex
\documentclass{article}

\usepackage[english]{babel}

\usepackage[letterpaper,top=2cm,bottom=2cm,left=3cm,right=3cm,marginparwidth=1.75cm]{geometry}

\usepackage{amsmath}
\usepackage[utf8]{inputenc}
\usepackage{graphicx}
\usepackage[colorlinks=true, allcolors=blue]{hyperref}
\usepackage{xcolor}
\usepackage{color, colortbl}
\definecolor{Gray}{gray}{0.9}
\usepackage{authblk}
\usepackage{eurosym}
\usepackage{float}
\floatstyle{plaintop}    
\usepackage{rotating}
\usepackage{caption}
\usepackage{multirow}
\usepackage[doublespacing]{setspace}
\usepackage{subfiles}
\usepackage{makecell}
\usepackage{natbib}
\setcitestyle{open={[},authoryear,close={]}}
\usepackage{caption}
\usepackage{changepage}
\usepackage{lineno}
\usepackage{forest}
\usepackage{tikz}
\usetikzlibrary{arrows.meta, quotes}
\usepackage{wasysym}

\usepackage{adjustbox}
\usepackage{graphicx}
\usepackage{booktabs,caption}
\usepackage[flushleft]{threeparttable}
\usepackage{soul}
\restylefloat{table}
\usepackage{booktabs}
\usepackage{dcolumn}
\newcolumntype{d}[1]{D..{#1}}

\begin{document}

\title{\Large {Human vs. Algorithmic Auditors: The Impact of Entity Type and Ambiguity on Human Dishonesty}}

\author{Marius Protte$^{*,\dagger}$ and Behnud Mir Djawadi$^{*}$}
\date{}

\maketitle

\begin{center}
  \noindent\textbf{Abstract}\\
    \noindent  
\end{center}

\singlespacing
{\small 
Human-machine interactions become increasingly pervasive in daily life and professional contexts, motivating research to examine how human behavior changes when individuals interact with machines rather than other humans. While most of the existing literature focused on human-machine interactions with algorithmic systems in advisory roles, research on human behavior in monitoring or verification processes that are conducted by automated systems remains largely absent. This is surprising given the growing implementation of algorithmic systems in institutions, particularly in tax enforcement and financial regulation, to help monitor and detect misreports, or in online labor platforms widely implementing algorithmic control to ensure that workers deliver high service quality. Our study examines how human dishonesty changes when detection of untrue statements is performed by machines versus humans, and how ambiguity in the verification process influences dishonest behavior. We design an incentivized laboratory experiment using a modified die-roll paradigm where participants privately observe a random draw and report the result, with higher reported numbers yielding greater monetary rewards. A probabilistic verification process introduces risk of detection and punishment, with treatments varying by verification entity (human vs. machine) and degree of ambiguity in the verification process (transparent vs. ambiguous). Our results show that under transparent verification rules, cheating magnitude does not significantly differ between human and machine auditors. However, under ambiguous conditions, cheating magnitude is significantly higher when machines verify participants' reports, reducing the prevalence of partial cheating while leading to behavioral polarization manifested as either complete honesty or maximal overreporting. The same applies when comparing reports to a machine entity under ambiguous and transparent verification rules. These findings emphasize the behavioral implications of algorithmic opacity in verification contexts. While machines can serve as effective and cost-efficient auditors under transparent conditions, their black box nature combined with ambiguous verification processes may unintentionally incentivize more severe dishonesty. These insights have practical implications for designing automated oversight systems in tax audits, compliance, and workplace monitoring.
}

\vspace{1cm}
\onehalfspacing

\onehalfspacing

\vspace{5mm}
{\noindent\rule{\textwidth}{0.1pt}\\
\noindent\footnotesize $^*$Paderborn University, Heinz-Nixdorf-Institute, Fürstenallee 11, 33102 Paderborn\\
\noindent $^\dagger$Corresponding author, \url{marius.protte@upb.de}\\
\noindent This research was funded by the Deutsche Forschungsgemeinschaft within the “SFB 901: On-The-Fly (OTF) Computing – Individualised IT-Services in Dynamic Markets” program (160364472).\\}

\clearpage

\doublespacing
\section{Introduction}

Human-machine interaction is ubiquitous in today’s world, driven by increasing automation and the growing reliance on algorithms and artificial intelligence (AI) in decision-making. AI, algorithmic advisors, and computerized decision support systems are employed in various domains, where they often outperform human judgment. Notable examples include medicine and healthcare \citep{cheng2016,gruber2019}, public administration \citep{kouziokasa2017,bignami2022}, autonomous driving \citep{levinson2011}, human resource management \citep{highhouse2008}, investment decisions \citep{tao2021}, insurance claim processing \citep{komperla2021}, tax audits \citep{black2022,baghdasaryan2022}, and criminal jurisdiction \citep{kleinberg2018}, among others. At the same time, demographic shifts and skilled labor shortages present pressing societal challenges, which are increasingly addressed through algorithmic and AI-based automation.

Despite algorithms often demonstrating superior predictive accuracy compared to human forecasters, people frequently prefer human input when given a choice between algorithmic and human forecasts \citep{dietvorst2015}. Likewise, individuals regularly disregard algorithmic advice in favor of their own judgment, even when doing so is not rational and leads to inferior outcomes \citep{burton2019,jussupow2020}. Conversely, the perceived reliability, consistency, and objectivity of algorithms can lead to over-reliance on their advice, particularly in structured and predictable tasks \citep{klingbeil2024,banker2019}. This duality in perception highlights the complexity of human attitudes toward machine-supported decision-making, as levels of algorithm acceptance and adherence typically vary widely across individuals and contexts \citep{fenneman2021}.

Many of the fields of application mentioned at the beginning inherently involve moral considerations to which individual differences in the perception of humans versus machines pertain. When algorithms act as ethical advisors, an asymmetry in their impact becomes apparent: algorithmic advice appears largely unsuccessful in promoting honest behavior, but is able to facilitate dishonest behavior \citep{leib2024}. Similarly, AI agents can function as enablers of unethical behavior in decisions that can be delegated by offering individuals a means to outsource or share the moral load imposed by unethical behavior \citep{köbis2021,bartling2012}. Regarding honesty, \citet{cohn2022} find significantly more cheating when individuals interact with machines than with humans, regardless of whether the machine has anthropomorphic features. Dishonest individuals actively prefer machine interaction when given an opportunity to cheat. Meanwhile, people cheat less in the presence of a robot \citep{petisca2022} or digital avatar \citep{mol2020} if it signals awareness of the situation than when being alone, even when it cannot intervene.

However, what happens to human dishonest behavior if machines can detect when someone lies or makes an untrue statement? Does behavior potentially change because of the machine entity itself or because of the ambiguity machines create through their "black box" nature? Concurrent with the tendency to use AI as advisors, algorithms are also used to monitor human conduct. For example, there is growing implementation of algorithmic systems in institutions, particularly in tax enforcement and financial regulation, to help monitor and detect misreports \citep[e.g.,][]{faundez2020use}. Similarly, online labor platforms widely implement algorithmic control to ensure that workers consistently deliver high quality
services \citep{wang2024does}. Despite the prevalence and impact of this form of human-machine interaction, we have limited understanding of how human dishonest behavior is shaped when their actions are subject to machine verification. We therefore ask the following research questions:

\vspace{5mm}

\textit{How does human dishonesty change when detection of untrue statements is performed by machines versus humans, and to what extent does ambiguity in the verification process influence dishonest behavior?}

\vspace{5mm}

We hereby make two important contributions. First, our research extends findings from the dishonesty literature by investigating scenarios where machines serve not as advisors or partners but as verification entities that detect untrue statements, an increasingly common human-machine interaction context. Second, while institutions such as tax authorities have increasingly implemented algorithmic systems to identify suspicious patterns in tax reports, our research clarifies whether the use of such machines creates a deterrence effect that reduces dishonesty. These insights may also provide valuable information for organizations implementing monitoring systems, where research regularly shows that electronic surveillance systems are often perceived negatively by employees and can even be associated with increased employee intentions to engage in counterproductive workplace behaviors.

To answer our research questions, we conduct an incentivized one-shot laboratory experiment that employs a modified version of the die-roll paradigm introduced by \citet{fischbacher2013}. Participants privately observe a random draw and report its outcome, with monetary payoffs tied to the reported number - creating an opportunity to profit from dishonesty. We introduce a two-stage verification process in which reports that may turn out to not coincide with the truth are sanctioned with a substantial monetary penalty. By incorporating elements of risk and uncertainty into the traditional dishonesty paradigm, our methodological approach maintains a generalizable framework that intentionally abstracts from domain-specific settings such as tax evasion or corruption. While these contexts share similar mechanisms of detecting and sanctioning deviant behavior, they frequently involve additional motivational factors such as civic duty, moral obligations, and imposing negative externalities on others that could confound the fundamental relationship between dishonest behavior and verification entity that we aim to isolate. We vary both the verification entity (Human vs. Machine) and the level of ambiguity involved in processing the die-roll reports (Black box vs. Transparent) to compare how participants' dishonest behavior is affected by who verifies their reports and how transparent the verification process is. We control for factors such as risk preferences, attitudes toward ethical dilemmas, perceived closeness to the auditor, and technology affinity.

The proceeding paper is structured as follows: 
Section \ref{hmd_literature_hypotheses} reviews prior research on perceptions of algorithmic entities, human dishonesty, and their intersection. With this context established, two hypotheses are derived for the experimental study. Subsequently, Section \ref{hmd_experiment} outlines the experimental design and procedure in detail.
Section \ref{hmd_results} presents descriptive results, followed by hypothesis testing and multivariate regression analysis.
Finally, Section \ref{hmd_discussion} offers an interpretation of the findings and concludes with a discussion of the study’s limitations and implications.

\section{Related Literature and Derivation of Hypotheses} \label{hmd_literature_hypotheses}

\subsection{Literature Overview}
\subsubsection{Algorithm perception}

Recent advances in human-machine interaction research increasingly focus on how individuals perceive algorithms and AI, particularly in the context of algorithm aversion and algorithm appreciation \citep[e.g.,][]{mahmud2022,jussupow2020,dietvorst2015,dietvorst2018,castelo2019,logg2019,fuchs2016}\footnote{Empirical research in this field can be broadly categorized into two strands: (1) studies in which humans interact with algorithms, programs, chatbots, or AI systems through a computer interface \citep[e.g.,][]{cohn2022,biener2021,dietvorst2015,logg2019}; and (2) studies involving humans interacting with anthropomorphic robots, focusing on perceived trustworthiness, intelligence, or reciprocity - often observed from a third-person perspective \citep[e.g.,][]{canning2014,ullman2014,sandoval2020}. The present study is concerned solely with the former type of interaction.}. Within this literature, the term \textit{algorithm} is often used as a broad synonym, encompassing various technological systems, including decision support systems, automated advisors, robo-advisors, digital agents, machine agents, forecasting tools, chatbots, expert systems, and AI-generated decisions \citep{mahmud2022}\footnote{From a technical standpoint, an algorithm is defined as a sequential logical process applied to a data set to accomplish a certain outcome. This process is automated and processes without human interference \citep{gillespie2016}.}. In line with this, we use the term "algorithm" to denote any technological system that applies a deterministic, stepwise process to decision-making \citep{dietvorst2020}.

Generally, attitudes toward algorithms vary widely among individuals. These attitudes are not fixed, but rather context-dependent, reflecting both algorithm aversion and algorithm appreciation \citep{fenneman2021,hou2021}.
\textit{Algorithm aversion} describes the tendency - whether conscious or unconscious - to resist relying on algorithms, even when they are demonstrably outperform human judgment. People frequently reject algorithmic advice in favor of their own or other humans’ opinions, despite being aware of the algorithm's superior accuracy and incurring material costs for doing so \citep{dietvorst2015,dietvorst2018,mahmud2022,jussupow2020}.
Although people frequently attribute near-perfect performance to algorithms \citep{dzindolet2002}, they are quicker to lose trust in them following errors, regardless of the error's context or severity \citep{renier2021}. In contrast, equivalent human mistakes are more readily excused \citep{madhavan2007}.
Conversely, \textit{algorithm appreciation} refers to situations in which individuals are more likely to follow identical advice when it originates from an algorithm rather than a human, often displaying greater confidence in such recommendations despite having little to no insight into the algorithm’s internal workings \citep{logg2019}. This effect is especially pronounced when the algorithm signals expertise \citep{hou2021}.
A systematic literature review by \citet{mahmud2022} concludes that algorithm acceptance varies along several demographic lines: older individuals and women tend to show greater aversion, while higher education is associated with greater acceptance. Moreover, algorithm aversion is often more pronounced among domain experts \citep{logg2019,jussupow2020}.

Both these directions of biased algorithm perception may result in economic inefficiencies. 
On the one hand, algorithms, despite not being entirely free of errors, consistently provide more accurate decisions than human counterparts \citep{dawes1989,logg2019}. Yet, in decisions under risk and uncertainty, individuals often disregard even high-quality algorithmic advice due to heightened sensitivity to potential errors, leading to suboptimal outcomes \citep{dietvorst2020,prahl2017,jussupow2020}. This reluctance is particularly evident in morally salient domains - such as medicine, criminal justice, or military contexts - where algorithmic input is frequently rejected even when it aligns with human decisions and produces efficient outcomes \citep{bigman2018}.
On the other hand, unreflective algorithm appreciation may results in over-reliance, where individuals defer to algorithmic recommendations despite contradictory contextual knowledge or better judgment. This can lead to suboptimal decisions with unintended consequences for both the decision-maker and affected third parties \citep{klingbeil2024}. For example, \citet{banker2019} find that consumers often rely to heavily on algorithmic recommendations, leading to inferior purchasing decisions. Similarly, \citet{krügel2022} demonstrate that individuals' decision-making in ethical dilemmas can be manipulated through overtrust in AI. 
Two key factors determining an individual’s unique degree of algorithm adherence, i.e., their inclination to either use or avoid algorithms, are anticipated efficacy and trust placed in the algorithmic system \citep{fenneman2021}. Perceived efficacy appears to have a stronger positive influence on willingness to rely on algorithms than discomfort or unease associated with using them \citep{castelo2019}. In terms of trust, similar factors as in human relationships - perceived competence, benevolence, comprehensibility, and responsiveness - also apply to automation. Additionally, perceptions specific to technology, such as reliability, validity, utility, and robustness, play an important role \citep{hoffman2013}.

\subsubsection{Human dishonesty}

People lie and cheat for their own benefit or for the benefit of others \citep{abeler2019,jacobsen2018}. However, despite being able to maximize their monetary payoffs, people often abstain from lying and cheating, for various reasons, e.g., general preferences for truth-telling, intrinsic lying costs, lying aversion, emotional discomfort and social image concerns \citep{abeler2014,abeler2019,bicchieri2009,khalmetski2019}. Additionally, lying behavior differs in magnitude, distinguishing between full liars (i.e., lying to the maximum extent possible), partial liars (i.e., exaggerating the actual outcome but not to the maximum), and fully honest individuals \citep{fischbacher2013,gneezy2018}.
Fittingly, previous experimental research (either in the lab or field) finds a considerable variance in cheating behavior among individuals with the opportunity to do so. Observed proportions of fully honest decision-making usually range between 40 \citep{fischbacher2013} and close to 70 percent \citep{peer2014,djawadi2015,gneezy2018}, while \citep{abeler2014} observe close to no cheating at all. The large-scale meta-study by \citet{gerlach2019} finds cheating rates of approximately 50\% across common experimental lying and cheating settings (sender-receiver games, die-roll tasks, matrix tasks). Meanwhile, similar heterogeneity can be found for the respective degree of dishonesty, as fractions of 2.5\% and 3.5\% lying to the maximum extent possible are observed by \citet{shalvi2011} and \citet{peer2014} respectively, while around 20\% of individuals lie to the maximum extent possible in \citet{fischbacher2013}. \citet{gneezy2018} find up to 47\% of subjects lying, and up to 91\% doing so to the maximum extent possible, depending on the combination of given reporting mechanism and having the opportunity to do so, as the degree of cheating generally appears to vary heavily with personal and situational factors \citep{gerlach2019}.

The possibility of lying and cheating in (nearly) all domains of human-machine interaction mentioned above imposes ethical challenges and financial costs to both businesses and society. \citet{cohn2022} find that individuals are more likely to engage in dishonest behavior when interacting with a machine rather than a human, regardless of whether the machine exhibits human-like characteristics. Moreover, individuals with an intention to cheat tend to prefer interacting with machines over humans. These patterns are largely attributed to diminished social image concerns and the perception that machines possess lower levels of agency \citep{cohn2022,biener2021}.

However, these findings stem from situations where untrue statements cannot be detected. In daily and economic life, such perfect concealment cannot always be guaranteed, and the recipient of a false statement might discover the truth. As machines may be perceived as more accurate than humans at detecting untrue statements, their presence as verifiers could potentially reduce cheating compared to human verification. Thus, the findings of the existing dishonesty literature may not apply to situations where detection is possible, necessitating empirical investigation of this specific context.

\subsection{Hypotheses}

Referring to the literature on algorithm aversion and appreciation, it becomes evident that in numerous daily and economic contexts, functionally equivalent actions performed by humans and machines can be differently perceived by human recipients. For the examination of detecting and potentially sanctioning dishonest behavior, there also exist competing arguments regarding whether dishonesty rates might increase or not when machines rather than humans verify the statements' truthfulness. On the one hand, algorithmic decisions are usually being perceived as more objective, consistent and less error-prone \citep{dzindolet2002,dzindolet2003,renier2021}. Human individuals intending to engage in dishonest behavior may therefore prefer human verification of their reports, anticipating a higher chance of avoiding detection and subsequent sanctions due to perceived limitations in human monitoring capabilities. Further, individuals may be more likely to act dishonestly when humans verify their statements because they believe humans exercise discretionary judgment based on empathy or fairness considerations. Such perceptions have been observed particularly in morally charged contexts \citep{dietvorst2015,mahmud2022,jauernig2022}. Machines, conversely, are conceptualized as rigid rule-followers lacking such affective capacities \citep{haslam2006,bigman2018,gogoll2018,niszczota2020}. On the other hand, empirical evidence indicates that human individuals perceive algorithmic surveillance more negatively than human surveillance \citep{schlund2024algorithmic}. Further, related literature provides indirect evidence that algorithmic monitoring does not prevent but in some cases even facilitate deviant behaviors. For instance, \citet{wang2024does} analyze data from a ride-hailing platform and finds that intensified algorithmic control implemented through work-related monitoring positively influences customer-directed deviant behavior among drivers. Similarly, \citet{liu2021digital} compare conventional taxi and Uber drivers, finding that despite enhanced algorithmic tracking capabilities in the latter context, route manipulation through detours that benefits drivers at passengers' expense is more prevalent in Uber rides compared to taxi rides during surge pricing periods. 
More direct evidence comes from experimental economics. \citet{cohn2022} find that individuals are significantly more likely to cheat machine agents than human ones, regardless of the medium (voice or text) or whether the machine features anthropomorphic traits. Dishonest individuals also show a preference for interacting with machines when given an opportunity to cheat. This behavior is attributed to social image concerns in interactions between humans, which have been previously identified as a key inhibitor of dishonesty \citep{abeler2019,khalmetski2019}. Similar findings are reported by \citet{biener2021}, who observe greater honesty when participants report the outcomes of unobserved, payoff-relevant random draws to a human rather than a chatbot. The degree of perceived agency, as well as considerations of social image and norms, appear to drive this difference. Social image concerns represent a plausible factor in our setting as well. Being detected and sanctioned by another human may carry higher reputational consequences for the individual than when detection occurs through algorithmic means, as machines are less likely to be perceived as forming judgments about character or moral worth. This asymmetry would suggest that algorithmic verification systems may inadvertently facilitate dishonest behavior by lowering the social costs that typically deter such conduct when human oversight is present. Given these competing arguments, we formulate our first hypothesis in a conservative manner without specifying the direction of potential behavioral differences:

\vspace{5mm}
\textbf{Hypothesis 1}: Human dishonest behavior will differ when their statements' truthfulness is verified by humans or machines.
\vspace{5mm}

As technological trends suggest that machines will increasingly be employed for automated detection processes, our second hypothesis focuses on machines as verification entities. Beyond psychological, biological, and ethical dimensions, perceptual differences between humans and machines are typically rooted in technological characteristics, where a central debate concerns whether algorithmic systems should operate through transparent rules or be deliberately kept ambiguous. There are indications that this discussion is also relevant for the human-machine interaction in our setting. As algorithms, by nature, tend to be opaque rather than transparent, they are frequently perceived as "black boxes" that convert some type of input into some type of output without revealing their internal logic \citep{tschider2020,mahmud2022}. Commonly, humans neither understand nor are aware of how algorithms function, which constitutes a major reason for them rejecting algorithms and their advice \citep{yeomans2019,dzindolet2002,kayande2009,mahmud2022}.
From the perspective of advice-taking, "opening the black box" through increasing transparency, accessibility, explainability, interactivity and tunability has been widely advocated to foster trust in and reduce aversion toward algorithms \citep{sharan2020,chander2018,holzinger2017,litterscheidt2020,shin2020}. However, it has been shown that even if an algorithm’s underlying logic is disclosed to the decision-maker, it may remain unintelligible, especially to non-experts \citep{onkal2009}.
Decision context also plays a crucial role. \citet{sutherland2016} find that humans are more inclined to rely on algorithms in uncertain environments. Contrastingly, \citet{longoni2019} report greater aversion to algorithmic decision-making in high-stakes environments rife with uncertainty such as healthcare. These mixed findings reflect a distinction in how humans perceive decisions under ambiguity (i.e., uncertainty) differently from decisions under risk, where potential outcomes and related probabilities are known \citep{ellsberg1961,einhorn1986,fox1995,chow2001}.
The influence of an algorithm’s black box nature specifically on human dishonest behavior is therefore not straightforward. 
Under transparent verification rules, dishonesty may reflect a rational cost-benefit analysis based on known probabilities, on the basis of which partial cheating nay reflect a rational outcome. Under ambiguity, however, where the likelihood of being detected and punished is unknown, such estimates become difficult. Therefore, transparency might actually encourage more dishonesty compared to an ambiguous detection process, as individuals can better assess these risks. 
In contrast, when detection probability parameters are unavailable, ambiguity may lead individuals to adopt an "all-or-nothing" strategy: either being fully honest to avoid any negative consequence or fully dishonest as uncertainty about detection applies equally to all untrue statements. In this vein, it is plausible to assume that if an individual decides to cheat under ambiguity they will do so more likely to the maximum extent possible. Whether the distribution under ambiguity consists of more honest than dishonest behavior is also not entirely clear. Literature has shown that ambiguity may intensify individual risk preferences \citep{gosh1997} and as most individuals are assumed to be risk-averse, this could result in a higher proportion of honest behavior. Conversely, ambiguity may also enable greater self-justification for dishonest behavior \citep[e.g.,][]{pittarello2015justifications}.

In summary, individual dishonest behavior is likely not only affected by the nature of the verification entity itself but also by whether machines operate the detection process under transparent or non-transparent rules. As there are convincing arguments for both more and less dishonest behavior under each rule type, we refrain from a directional prediction in formulating our second hypothesis:

\vspace{5mm}
\textbf{Hypothesis 2}: Human dishonest behavior will differ when their statements' truthfulness is verified by machines under transparent or undisclosed rules.\par


\section{Experiment} \label{hmd_experiment}

We conducted a one-shot, incentivized laboratory experiment in which participants entered a prize draw with a potential payoff of up to \euro{90}. The final payoff depended on each participant’s decision and the outcomes of up to two lotteries. Only one winner was drawn per session, in line with a random incentive system - a well-established approach in experimental economics that has been shown to produce similar behavior as under deterministic payoff schemes \citep{charness2016,camerer1999,bolle1990,tversky1981}.

\subsection{Experimental design} \label{hmd_design}

The experiment comprised two main parts: the \textbf{Choice Part} and the \textbf{Verification Part}.

\begin{figure}[H]
  \includegraphics[scale=0.4]{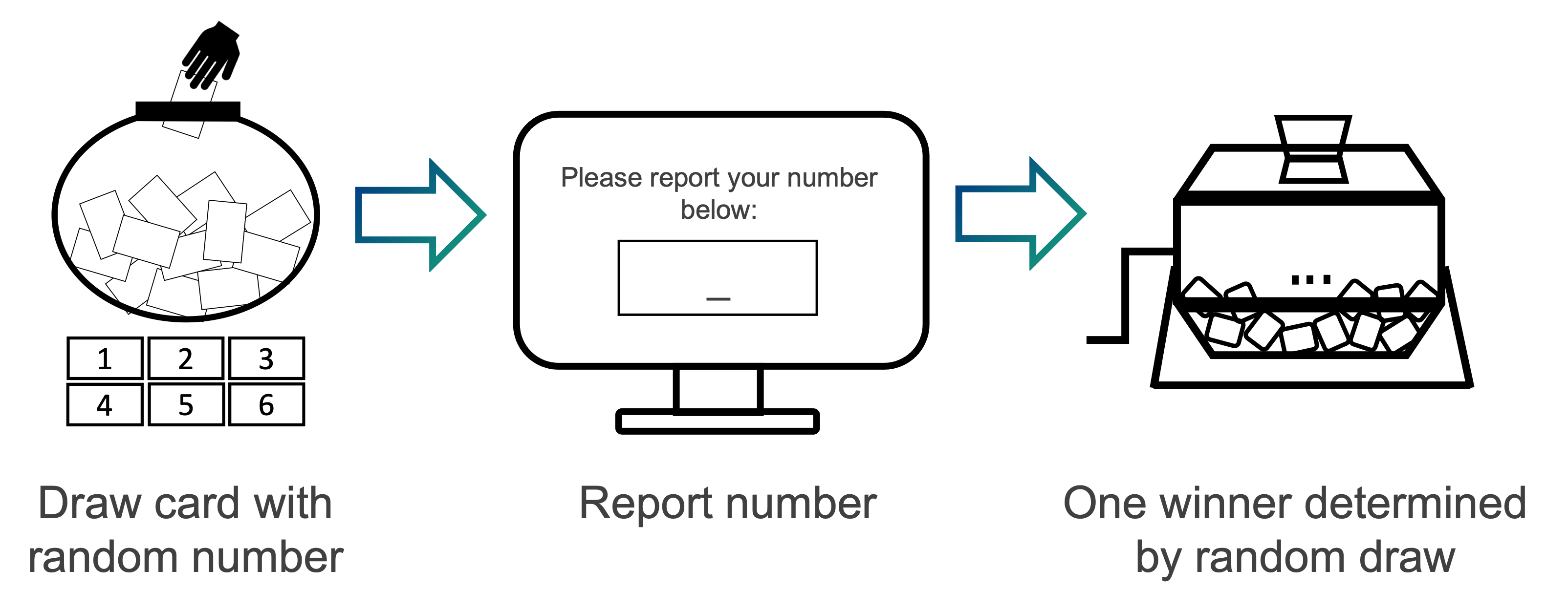}
  \centering
  \caption{Overview of the Choice Part of the experiment (all participants)}
  \label{fig:hmd_choicepart}
\end{figure}

In the \textbf{Choice Part}, illustrated in Figure \ref{fig:hmd_choicepart}, subjects drew exactly one card randomly from an urn containing 100 cards numbered between 1 and 6. Subsequently, they confidentially reported their drawn number via a computer interface. Importantly, the reported number - in conjunction with Verification Part results - would later determine the prize payoff for one randomly selected winner, calculated as the reported number multiplied by \euro{15} (payoff range: \euro{15} to \euro{90}). This setup created the opportunity for subjects to increase their potential payoff by overreporting the drawn number.
After submitting their report, participants completed a series of questionnaires (see Section \ref{hmd_exprocedure}), before the prize winner was determined.

In the \textbf{Verification Part}, the winner underwent a verification procedure comprising up to two lotteries:

\begin{itemize}
    \item In \textit{Lottery 1}, a number between 1 and 10 was randomly drawn. If this number was greater than the participant’s reported number, no additional check occurred, and the full payoff (reported number × \euro{15}) was paid. If the number was less than or equal to the reported number, the participant's actual drawn card was checked.
    \item In the check, if the reported and actual numbers matched, the prize winner received the full payoff. 
    \item If they mismatched, \textit{Lottery 2} was triggered: An urn containing numbers from 1 up to the reported number was used to randomly draw one number. If this drawn number was less than or equal to the participant's actual number, the price winner still received the full payoff. Otherwise, the payoff was reduced to the actual number multiplied by \euro{7.50} (payoff range: \euro{7.50} to \euro{37.50}).
\end{itemize}

Thus, the verification procedure incorporated two central design features. First, the probability of a card check increased with the magnitude of the reported number - similar to materiality thresholds in accounting, where more conspicuous reports are subject to greater scrutiny. Second, the probability of punishment, conditional on being checked, increased with the discrepancy between the reported and actual number. This mechanism allowed subjects to potentially receive the full payoff despite overreporting, thereby mimicking discretionary tolerance in real-world verifications, where minor deviations may be overlooked while larger discrepancies are more likely to result in sanctions.

\begin{figure}[H]
  \includegraphics[scale=0.25]{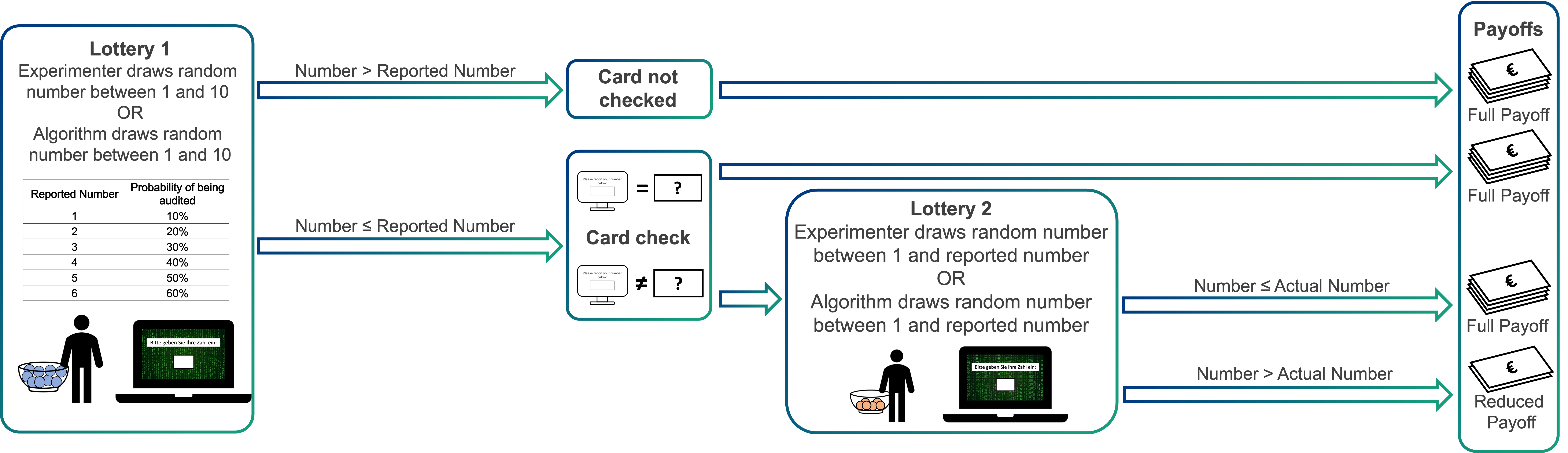}
  \centering
  \caption{Overview of the Verification Part of the experiment (only prize draw winner)}
  \label{fig:hmd_auditpart}
\end{figure}

The structure of the Verification Part is illustrated in Figure \ref{fig:hmd_auditpart}.

Theoretically, for a risk-neutral decision-maker, the payoff-maximizing strategy is to always report a 6, regardless of the actual number drawn. The design of Lottery 2 ensured that cheating of equal magnitude carried identical punishment probabilities - for instance, overstating a 3 as a 5 was punished with the same likelihood as overstating a 2 as a 4. Meanwhile, the likelihood of punishment increased with the extent of the misreport: for example, if two subjects both reported a 6, the subject who actually drew a 2 faced a higher probability of being penalized than one who drew a 5. 
The formal derivation of this utility function is provided in Appendix \ref{hmd_derivation}.

\textbf{Treatment Conditions}

We implemented four experimental treatments: \textbf{Human (H), \textbf{Machine (M)}, \textbf{Human Black Box (HB)}, and \textbf{Machine Black Box (MB)}}. 
In the Human treatment, the verification process was conducted by a human agent (i.e., the experimenter), whereas in the Machine treatment, it was executed by a computerized, rule-based algorithm. To emphasize the verification entity's role, both were visually represented using pictographs in the instructions (see Figure \ref{fig:hmd_human_audit} in Appendix \ref{hmd_audit}). Procedures in the black box versions (HB and MB) mirrored their respective non-black box treatments (H and M), except that the verification rules were not disclosed to participants.

In the Human treatments, the lotteries of the verification part were physically implemented using numbered balls drawn from urns. In the Machine treatments, the process was simulated by a computer algorithm, with visual feedback (e.g., animations; see Figures \ref{fig:hmd_algo_audit_example_input} to \ref{fig:hmd_algo_audit_example_tolerance} in Appendix \ref{hmd_audit}) provided to convey the impression of data processing. Critically, the underlying verification rules and their associated probabilities were held constant across all treatments; only the entity who conducted the process (human vs. machine) varied. 

In the black box conditions, the exact same procedures were applied (verification rules and probabilities remained identical). However, subjects were only informed that a human or machine would decide whether a card check would occur and, in the case of a mismatch, whether the payoff would be reduced. To reflect this lack of procedural transparency, the verification steps were referred to as "Decision 1" and "Decision 2" in the instructions. 

In all treatments, while participants were informed the urn contained numbers 1 to 6, they were not told the actual distribution. The true composition of the urn was 95 cards displaying the number 2, while the numbers 1, 3, 4, 5, and 6 were each represented by a single card. This design ensured that most participants would draw a 2, allowing for individual-level analysis of dishonest behavior and increasing opportunities for overreporting. It would also largely prevent reduction of the sample size for the analysis due to subjects drawing a 6, which left them no opportunity to be dishonest. 
After each session, the remaining cards in the urn were counted to infer the actual distribution of numbers drawn. If all five non-2 cards remained, any report higher than 2 could be clearly identified as dishonest. If one or more of the five non-2 cards had been drawn, one observation with a report of a 6 would be randomly excluded from the dataset per card drawn, to obtain a conservative estimate of dishonest behavior. 

This approach did not disadvantage any participant, as the distribution of cards was not disclosed in the instructions.
The decision to equip the urn with a majority of cards numbered with a "2" instead of a "1" was made to avoid triggering "revenge cheating" (i.e., retaliation due to receiving the lowest possible draw) and to ensure participants faced a meaningful trade-off between honesty and financial gain. By drawing a "2" with the highest probability, truthful reporting would yield a \euro{30} payoff for the prize winner, which is already substantial for an experiment participation of around 45-minutes, but could potentially be tripled through dishonest reporting.

\subsection{Experimental procedure} \label{hmd_exprocedure}

The experiment was conducted in December 2023 at the Business and Economic Research Laboratory (BaER-Lab, \url{www.baer-lab.org}) at Paderborn University and computerized using oTree \citep{chen2016}. Subjects were
recruited via the online recruiting system ORSEE \citep{greiner2015} and were only allowed to participate in one session. In total, ten sessions were run (Human: 3, Machine: 3, Human Black Box: 2, Machine Black Box: 2). Each session lasted 30-45 minutes.

Participants were randomly assigned to individual computer workplaces in cubicles to ensure privacy and were instructed not to communicate during the session. After receiving written instructions (see Appendix \ref{hmd_instructions}) and being given time to read them carefully, participants completed extensive comprehension checks to ensure a sufficient understanding of the experimental rules and payoff conditions. They could only proceed after answering all questions correctly. Consequently, subjects were, at least implicitly, aware of the opportunity to misreport before making any decisions in the experiment.

The Choice Part began once all subjects had successfully completed the comprehension checks. The experimenter moved from cubicle to cubicle, presenting an urn containing the number cards to each subject. After the drawing process was completed, the experiment automatically advanced to the reporting screen, where subjects entered their reported number. To encourage thoughtful decision-making, participants were not subjected to any time limit.

After confirming their choice, subjects completed a series of questionnaires (see Appendix \ref{hmd_questionnaire}). First, they were asked whether they generally preferred a human or a machine to perform the verification process. Second, subjects were asked which of the two entities they generally perceived as more error-prone and which as having greater discretion. 
Subsequently, subjects answered standardized questionnaires on affinity for technology interaction \citep{franke2018}, attitudes toward ethical dilemmas \citep[adapted from][]{blais2006}, a pictorial measure of interpersonal closeness (adapted for inter-entity comparison) \citep[][based on \citet{aron1992}]{schubert2002}, the general risk preference measure by \citet{dohmen2011}, as well as demographic questions.

Once all questionnaires were completed, one prize winner was randomly selected using the cubicle numbers. Non-winning participants received a fixed payment of \euro{7.50} in cash to compensate for their participation time\footnote{This is three times the amount of the laboratory's usual show-up fee in experiments with individual performance-dependent incentives.} and were then dismissed. 

The Verification Part was conducted privately with the winner to preserve anonymity and minimize social influence \citep{bolton2021}\footnote{While social image concerns toward the experimenter cannot be ruled out entirely, comparative statics ensure interpretability of treatment differences between groups.}. The two lotteries were implemented based on the entity type of the respective treatment, following the procedure described in Section \ref{hmd_design}. The winner received their (full or reduced) payoff in cash, concluding the session.

\section{Results} \label{hmd_results}

In total, one-hundred-seventy ($N=170$) student subjects participated in the experiment. Of these, 48 were randomly assigned to the Human treatment (H), 41 to the Machine treatment (M), 43 to the Human Black Box treatment (HB), and 38 to the Machine Black Box treatment (MB) respectively. In the analysis, each subject constitutes one independent observation in the analysis. An overview of demographic characteristics is provided in Table \ref{tab:hmd_demographics}. Participants were, on average, 22 years old, with ages ranging from 18 to 36. Women constituted 56\% of the sample, and gender distribution did not differ significantly between treatments (Pearson $\chi^2(3) = 0.43, p = 0.935$). Multiple fields of study were represented, with Business Administration \& Economics (56.5\%) being the most common. The distribution of of fields of study did not differ significantly between treatments (Pearson $\chi^2(6) = 11.14, p = 0.084$). 

\begin{table}[H]
    \centering
    \caption{Demographic statistics}
    \begin{tabular}{ l l c c c c c }
         \hline
         \hline
          & & \textbf{H} & \textbf{M} & \textbf{HB} & \textbf{MB} & \textbf{Overall} \\ \cline{3-7}
         \multicolumn{2}{l}{\textbf{Number of observations}} & 48 & 41 & 43 & 38 & 170\\
         \hline
         \multicolumn{2}{l}{\textbf{Age}} & & & & & \\
          & Mean & 21.8 &  21.8  & 21.9 & 22.5 & 22.0 \\
          & Std. deviation & 3.2 & 3.5 & 3.5 & 4.1 & 3.5\\
         \hline
         \multicolumn{2}{l}{\textbf{Gender (\%)}} & & & & &  \\
          & Female & 54.2  & 58.5 & 58.1 & 52.6 & 55.9 \\
         \hline
         \multicolumn{2}{l}{\textbf{Field of studies (\%)}} & & & & & \\
          & Business Administration \& Economics & 56.3  & 68.3  & 58.1  & 42.1  & 56.5 \\
          & Cultural Sciences & 37.5 & 22.0 & 37.2 & 36.8 & 33.5\\
          & Natural Sciences & 6.3 & 9.8 & 4.7 & 21.1 & 10.0 \\
         \hline
         \hline
    \end{tabular}
    \label{tab:hmd_demographics}
\end{table}

\subsection{Dishonest behavior}

Similarly to \citet{djawadi2015}, our design enables a direct and relatively precise measurement of dishonest behavior - in contrast to prior experimental studies that infer dishonesty by comparing reported outcomes to theoretical distributions \citep[see e.g.,][]{abeler2014,hao2017,fischbacher2013,shalvi2011,jacobsen2016} - by comparing the distribution of numbers drawn with the distribution of numbers reported. We use two dependent variables to measure cheating behavior: frequency and magnitude of overreporting, with the primary focus on the latter. 

\begin{figure}[H]
  \includegraphics[scale=0.55]{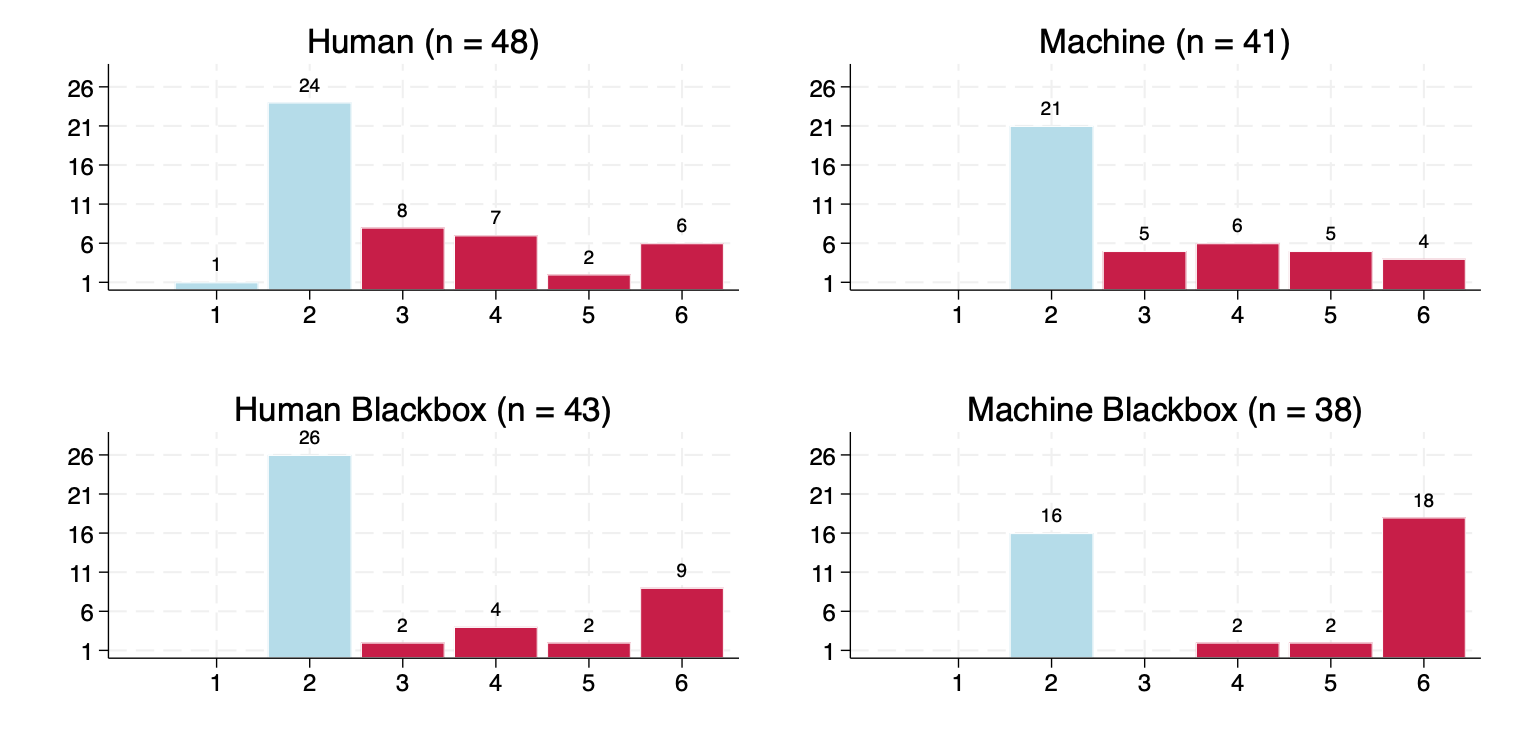}
  \centering
  \caption{Frequency Distributions of Reported Numbers, by Treatment}
  \label{fig:hmd_report}
\end{figure}

Figure \ref{fig:hmd_report} displays the frequency distributions of reported numbers by treatment. On average, subjects in the Human, Machine, and Human Black Box treatments reported numbers close to 3 (H: 3.06, M: 3.17, HB: 3.21), while subjects in the Machine Black Box treatment reported an average of 4.16. Reporting distributions differ significantly between groups (Pearson $\chi^2(15) = 33.07, p = 0.005$).

In all treatments except the Human condition, no other numbers than 2 were drawn. In the Human treatment, the number 1 was drawn and accurately reported. Therefore, no exclusions of observations from the reported distributions were necessary, and any reported number above 2 can be interpreted directly as cheating. 

In the non-black box groups, nearly half of the participants overreported: 23 out of 48 (47.9\%) in the Human treatment and 20 out of 41 (48.7\%) in the Machine treatment reported a higher number than they actually drew. Overreporting was less prevalent in the Human Black Box group (17 out of 43, or 39.5\%), while the highest rate occurred in the Machine Black Box group (22 out of 38, or 57.9\%). However, these differences in reporting rates are not statistically significant (Pearson $\chi^2(3) = 2.73, p = 0.435$).

\begin{table}[H]
    \centering
    \caption{Summary statistics of cheating behavior by treatment}
    \label{tab:hmd_summary}
    \begin{threeparttable}
    \begin{tabular}{ l l c c c c c }
         \hline
         \hline
         & & \textbf{H} & \textbf{M} & \textbf{HB} & \textbf{MB} & \textbf{Overall} \\ 
         \hline
         \multicolumn{2}{l}{\textbf{Type of behavior (\%)}} &  &  &  & \\ 
         & Honest & 52.1 & 51.2 & 60.5 & 42.1 & 51.7 \\ 
         & Partial cheating & 35.4 & 39.0 & 18.6 & 10.5 & 26.5 \\ 
         & Full cheating & 12.5 & 9.7 & 20.9 & 47.4 & 21.8 \\ 
         \hline
         \multicolumn{2}{l}{\textbf{Magnitude of cheating}} &  &  &  & \\ 
         & Mean & 2.26 & 2.40 & 3.06 & 3.73 & 2.85 \\ 
         & Median & 2 & 2 & 4 & 4 & 3 \\ 
         & Std. Deviation & 1.21 & 1.10 & 1.14 & 0.63 & 1.19 \\ 
         \hline
         \hline
    \end{tabular}
    \begin{tablenotes}
    \small
      \item \textit{Note}: Summary statistics of behavior type (relative frequencies) and cheating magnitude (among cheaters; absolute magnitude) by treatment. Instances of dishonest reporting: H: n = 23; M: n = 20; HM: n = 17; MB: n = 22.
    \end{tablenotes}
    \end{threeparttable}
\end{table}

Following conventions in related studies, we classify participants who overreported to the maximum extent possible (i.e., reporting a "6") as full cheaters, and those who overreported by a smaller margin as partial cheaters. Overall, the distribution of honest participants, partial cheaters, and full cheaters (see Table \ref{tab:hmd_summary}) differs significantly between treatments (Pearson $\chi^2(6) = 25.93, p < 0.0001$). In the non-black box groups, partial cheaters outnumber full cheaters. In the Human Black Box group, the proportions are roughly equal. In contrast, the Machine Black Box condition shows a substantially larger share of full cheaters, with partial cheaters being nearly absent. Notably, over half of participants were honest in the Human, Machine, and Human Black Box groups respectively, while the number of subjects who overreported to the maximum extent in the Machine Black Box group was higher than the number of honest subjects.

Regarding the magnitude of cheating, among cheaters, the average overreporting exceeded two numbers in all conditions, but was markedly higher in the black box groups. Consistently, the median magnitude of cheating was 2 in the non-black box treatments and 4 in the black box treatments. 
A Kruskal–Wallis equality-of-populations rank test with ties reveals a statistically significant difference in cheating magnitude across groups (Pearson $\chi^2(15) = 21.64, p = 0.0001$). The Machine Black Box group not only shows the highest average cheating magnitude but also the lowest standard deviation, indicating more consistent and extreme overreporting, reflecting the group with the highest proportion of full liars.

Comparing magnitudes of cheating under transparent verification rules, we find no significant differences between the Human and Machine entity treatments (Mann–Whitney U-test: $|z| = 0.48, p = 0.6357$), as the average magnitude of cheating is only marginally higher in the Machine treatment than in the Human treatment. 
Under undisclosed rules, however, we observe a notable difference in cheating magnitude, as average overreporting is 0.7 higher in the Machine Black Box group than in the Human Black Box group - a difference that is statistically significant (Mann-Whitney U-test: $ |z| = 2.09, p = 0.0442$). 
We therefore find partial support for \textbf{Hypothesis 1}, as the average magnitude of cheating differs by verification entity, but only under undisclosed verification rules.

Focusing on the machine groups under transparent and undisclosed verification rules, we observe a substantial increase in the average extent of overreporting - by approximately 1.3 - with the introduction of ambiguity about verification rules in the Machine Black Box group compared to the Machine group. The difference is highly statistically significant (Mann-Whitney U-test: $|z| = 4.03, p < 0.0001$). 
Therefore, we find support for \textbf{Hypothesis 2}: average magnitude of cheating toward a machine as verification entity differs between transparent and undisclosed processing rules, as ambiguity appears to lead to a higher magnitude of cheating.
For comparison, overreporting toward a human as verification entity significantly increased by, on average, 0.8 from the Human to the Human Black Box (Mann-Whitney U-test: $|z| = 2.02, p = 0.0469$) \footnote{We conducted hypothesis testing based on the sub-sample of individuals who engaged in dishonest behavior, i.e., overreported their drawn number, as we argue that including honest reports would dilute the true extent of damage caused by cheating. Naturally the average magnitude of overreporting declines when these are incorporated (H: 1.1; M: 1.2; HB: 1.2; MB: 2.2). Nevertheless, key statistical results would remain robust: under undisclosed verification rules, the entity effect remains statistically significant (Mann–Whitney U-test: $|z| = 2.19, p = 0.0296$), as does the effect of ambiguity with a machine verifying the reports (Mann–Whitney U-test: $|z| = 2.28, p = 0.0214$), while still no significant difference is observed between entities under transparent rules (Mann–Whitney U-test: $|z| = 0.25, p = 0.8076$).}.

To compare effect sizes, we calculate Cohen’s d with bootstrapped standard errors (see Figure \ref{tab:hmd_effsize_power} in Appendix \ref{hmd_appendix}). The entity effect is negligible in size under transparent verification rules ($d = -0.12$), while increasing to $d = -0.75$ under ambiguous rules, which can be classified as medium to large based on conventional benchmarks \citep{cohen1988}. Analogously, the effect of ambiguity in machine verification can be considered (very) large ($d = -1.50$).

\subsection{Control variables}

\begin{figure}[H]
  \centering
  \begin{minipage}[b]{0.45\textwidth}
    \includegraphics[width=\textwidth]{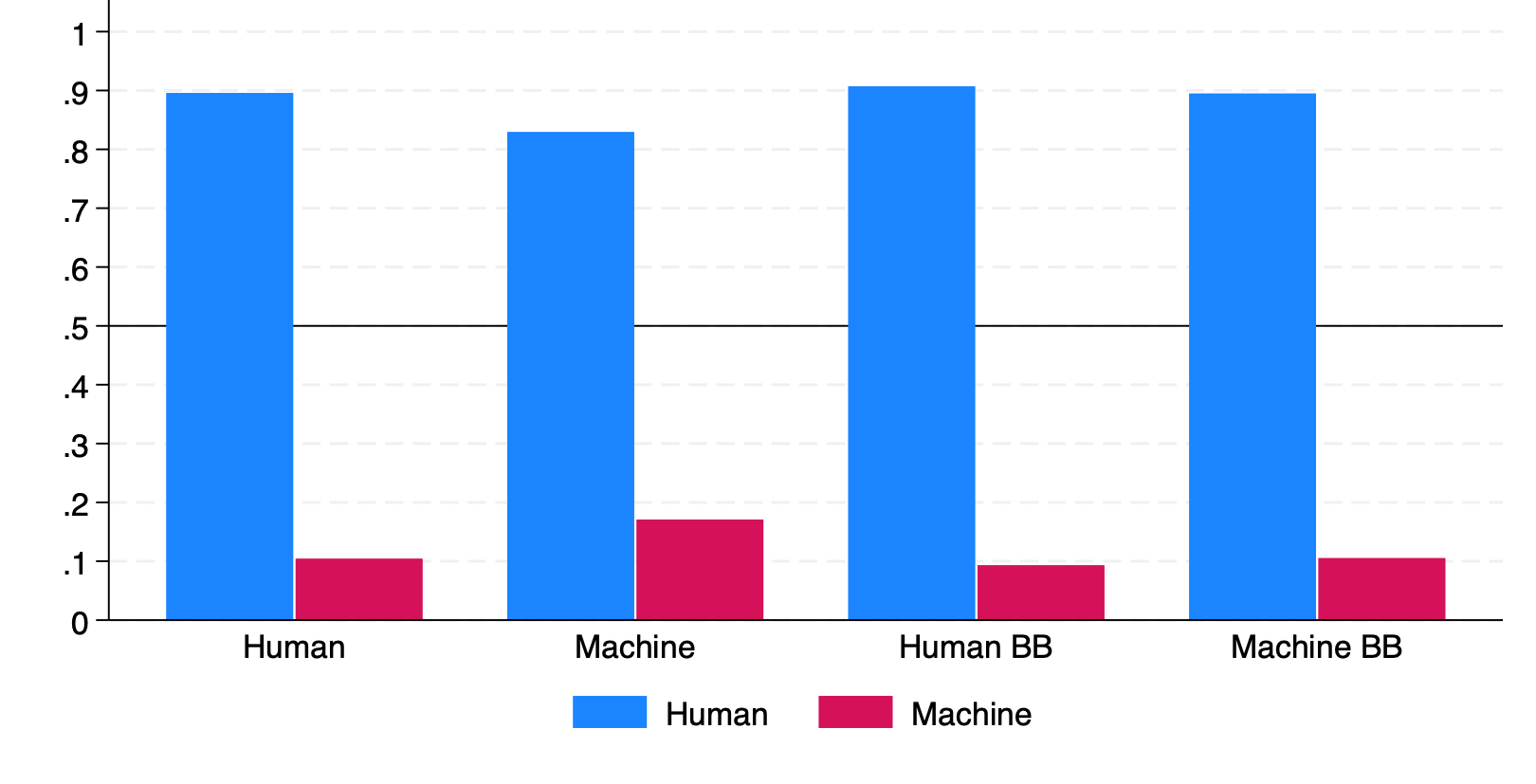}
    \caption{\footnotesize Perceived Error-proneness, by Treatment}
    \label{fig:hmd_errorexp}
  \end{minipage}
  \hfill
  \begin{minipage}[b]{0.45\textwidth}
    \includegraphics[width=\textwidth]{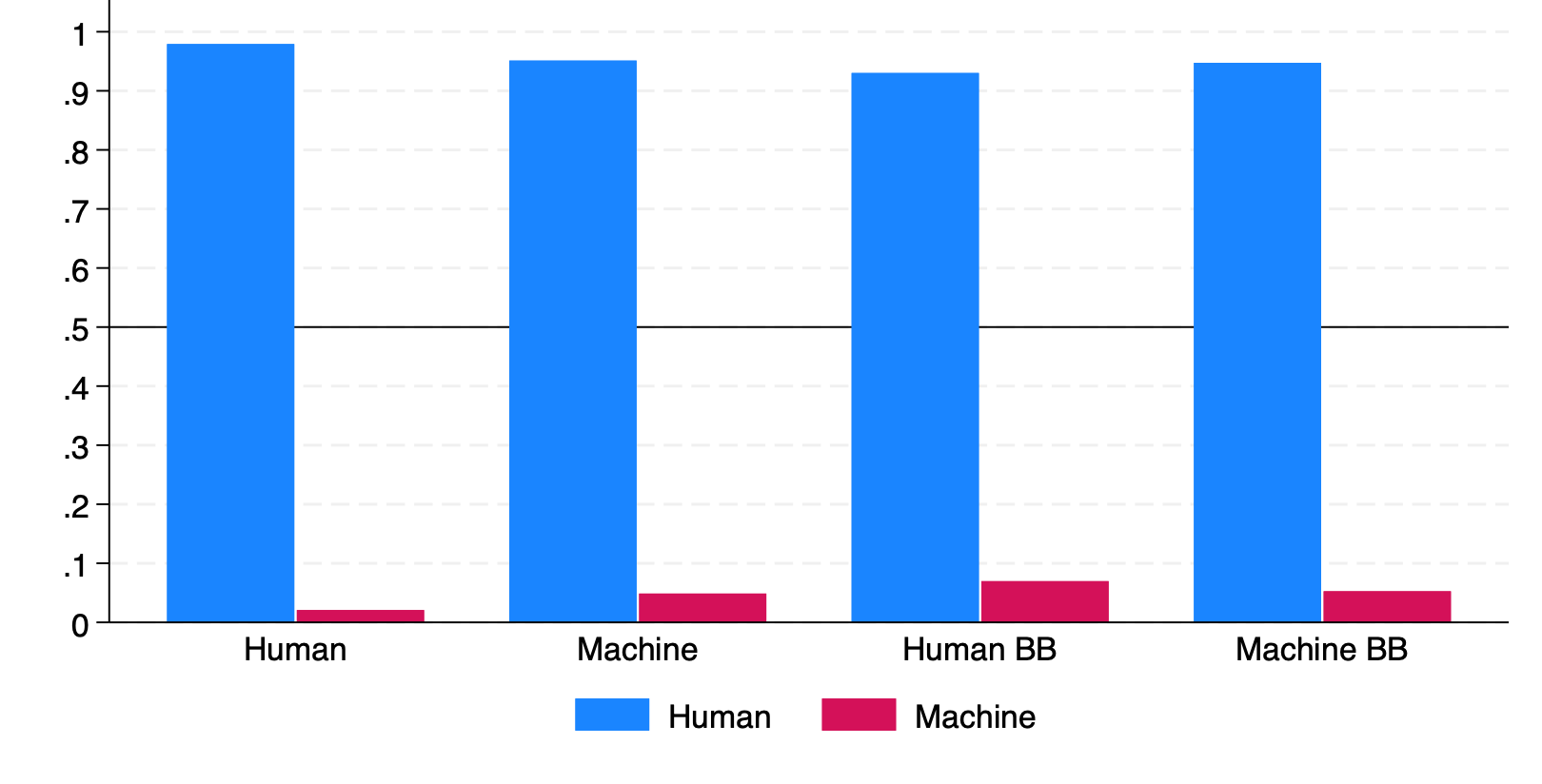}
    \caption{\footnotesize Perceived Discretion, by Treatment}
    \label{fig:hmd_discretion}
  \end{minipage}
\end{figure}

The analysis of our questionnaire data provides strong support for the assumption that participants perceive humans as both more error-prone and more discretionary in their decision-making, as illustrated in Figures \ref{fig:hmd_errorexp} and \ref{fig:hmd_discretion}, Binomial tests for both variables yield results significantly different from 0.5 - which would indicate indifference - across all four treatment groups ($p < 0.0000$). Moreover, response distributions do not differ significantly between groups (error-proneness: Pearson $\chi^2(3) = 1.50, p = 0.681$; discretion: Pearson $\chi^2(3) = 1.26, p = 0.739$).

\begin{figure}[H]
  \includegraphics[scale=0.35]{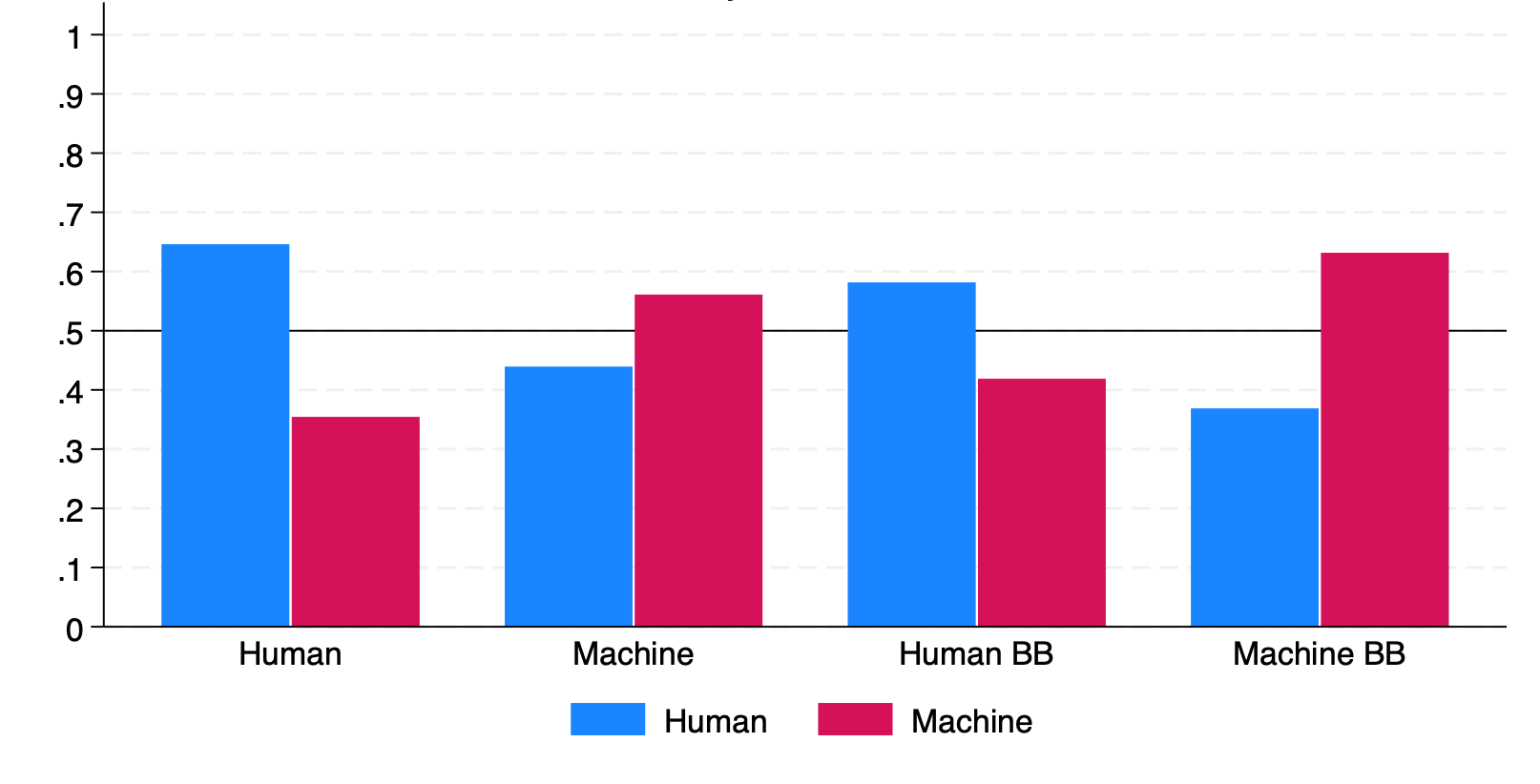}
  \centering
  \caption{Stated Preference for Verification Entity, by Treatment}
  \label{fig:hmd_preference}
\end{figure}

Findings are less conclusive regarding participants’ preferred entity for verifying the reports (see Figure \ref{fig:hmd_preference}). In both human treatment groups, participants tended to prefer a human as verification entity, whereas in the machine treatments, preferences leaned toward a machine as verification entity. However, in none of the groups did the distribution of preferences differ significantly from an even 50/50 split (see Table \ref{tab:hmd_controls} in Appendix \ref{hmd_appendix} for Binomial test results by group). The apparent tendency to prefer the respective verification entity encountered during the experiment may reflect a default option effect \citep{johnson2003}, as preferences were elicited post-experiment.

Furthermore, standardized questionnaire controls indicate that self-reported affinity for technology interaction, sensitivity to ethical dilemmas, perceived closeness to the verification entity, and stated risk preferences did not differ substantially across experimental groups as shown in Table \ref{tab:hmd_questionnaire_controls}\footnote{For pairwise treatment comparisons of cheating frequency and control variables see Table \ref{tab:hmd_pairwise} in Appendix \ref{hmd_appendix}}.

\begin{table}[H]
    \caption{Summary Statistics and Between-Group Comparison of Questionnaire Items}
    \label{tab:hmd_questionnaire_controls}
    \begin{threeparttable}
    \begin{tabular}{l c c c c c c c c}
         \hline
         \hline
          & \textbf{H} & \textbf{M} & \textbf{HB} & \textbf{MB} & \textbf{Total} & & \multicolumn{2}{c}{Kruskal-Wallis-H}\\ \cline{2-6} \cline{8-9}
        Number of observations & 48 & 41 & 43 & 38 & 170 & & $\chi^2(3)$ & $p$\\
        \hline
        & & & & & & & & \\        
        Affinity to technology interaction & 3.58 & 3.41 & 3.62 & 3.92 & 3.62 & & 5.16 & 0.161\\
                     & (0.87) & (0.84) & (.84)  & (1.23) & (0.93) &  & &\\
        & & & & & & & & \\        
        Ethical dilemma sensitivity & 4.15 &  4.19 & 4.26 & 4.10 & 4.18 & & 2.02 & 0.569\\
                     & (.47) & (.41) & (.47) & (.58) & (.48) &  & &\\      
        & & & & & & & & \\        
        Interpersonal closeness & 2.71 & 3.02 & 2.84 & 3.00 & 2.88 & & 5.06 & 0.167 \\
                     & (1.46) & (1.15) & (1.54) & (1.23) & (1.36) & & &\\ 
        & & & & & & & & \\        
        Risk preferences & 5.77 & 6.22 & 5.77 & 6.03 & 5.94 & & 1.37 & 0.713 \\
                     & (2.15) & (2.24) & (2.16) & (2.11) & (2.15) & & & \\ 
         \hline
         \hline
    \end{tabular}
    \begin{tablenotes}
    \small
      \item \textit{Note}: Summary statistics for affinity to technology interaction (6-point scale), sensitivity towards ethical dilemmas (5-point scale), perceived closeness towards the verification entity (7-point scale), and self-reported risk preferences (11-point scale). Standard deviations are reported in parenthesis. Kruskal-Wallis-H reports p-values for Kruskal-Wallis H-tests with ties between experimental groups.
    \end{tablenotes}
    \end{threeparttable}
\end{table}

Across all subjects, those who overreported and thus cheated reported a significantly higher willingness to take risks (Mann–Whitney U-test: $|z| = 2.62, p = 0.0085$). On average, cheaters indicated a general risk tendency of 6.4 (median: 7) on an 11-point scale, compared to 5.5 (median: 5.5) among honest participants. Also, the willingness to take risks was significantly positively correlated with the magnitude of cheating (Spearman's $\rho = 0.355, p = 0.0011$). 

Also, gender differences were evident: women cheated significantly less frequently than men (Pearson $\chi^2(1) = 13.36, p < 0.0001$), with 35.8\% of female and 64.0\% of male participants overstating their drawn number. However, the magnitude of cheating did not differ significantly between genders (Mann–Whitney U-test:  $|z| = 0.67, p = 0.5032$). 

The other demographic and control variables did not differ significantly between honest and dishonest participants, nor were they significantly associated with the extent of cheating (see Table \ref{tab:hmd_by_cheater} in Appendix \ref{hmd_appendix}).

\subsection{Regression analysis}

In addition to our non-parametric analysis, we conduct multivariate regression analysis to gain a deeper understanding of the relationship between cheating behavior and its potential determinants. 
Based on the sub-sample of individuals who cheated ($n = 82$), we examined the factors influencing the extent to which participants overstated their drawn number. Specifically, we regressed the magnitude of cheating on the type of verification entity and the ambiguity level of verification rules, along with demographic, control, and entity-perception variables. Table \ref{hmd_regression_magnitude} presents the results of the multivariate OLS regression, comparing multiple model specifications. 

The baseline model (Column 1) includes only treatment indicators as independent variables, while subsequent models add demographic variables (Column 2), control variables (Column 3), and dummy variables indicating matches between the assigned verification entity and participants’ stated entity preferences, perceptions of error-proneness, and perceived discretion (Column 4) respectively. All available variables are included in the full model (Column 5). For the sake of completeness, Tables \ref{hmd_regression_frequency} and \ref{hmd_logit_likelihood_margins} in Appendix \ref{hmd_appendix} present a linear probability model and marginal effects from a logistic regression estimating the independent variables' influence on the likelihood of cheating across the full sample. Both used the same model specifications as those employed in the regression for cheating magnitude. These robustness checks yield results consistent with our non-parametric analysis, with gender and general risk preferences emerging as the only statistically significant and substantively meaningful predictors of likelihood to cheat. For instance, being female is associated with a 30.7 percentage-point lower probability of overreporting.

\begin{table}[H]
\centering 
\caption{OLS Regression for Magnitude of Cheating} 
\label{hmd_regression_magnitude}
\begin{threeparttable}
\begin{tabular}{l l c c c c c}
\hline
\hline
 & & \multicolumn{5}{c}{Dependent variable: Magnitude of cheating} \\ \cmidrule{3-7} 
 & & \multicolumn{1}{c}{(1)} & \multicolumn{1}{c}{(2)} & \multicolumn{1}{c}{(3)} & \multicolumn{1}{c}{(4)} & \multicolumn{1}{c}{(5)} \\ 
\midrule
 \multicolumn{2}{l}{Intercept} & $2.261^{***}$ & $ 1.480^{*}$ & -1.069  & $2.218^{***}$ & -0.835\\ 
  & & (0.253) & (0.675) & (1.068)  & (0.616) & (1.564) \\
  &  & & & &  &  \\
 \multicolumn{2}{l}{Treatment} &  &  &  &  &  \\ 
  & \textit{Machine} & 0.139 & 0.067 & $2.500^{**}$  & 0.245 & 1.823 \\ 
  & &(0.353) & (0.371) & (0.795)  & (0.615) & (0.924) \\ 
  & \textit{Human Black Box} & $0.798^{*}$ & 0.625 & 0.478  & $0.816^{*}$ & 0.486 \\ 
  & & (0.375) & (0.364) & (0.364) & (0.364) &  (0.352) \\ 
  & \textit{Machine Black Box} & $1.466^{***}$ & $1.640^{***}$ & $3.957^{***}$ & $1.600^{**}$ & $3.520^{***}$ \\ 
  & & (0.287) & (0.252) & (0.814) & (0.580) & (0.944) \\ 
  &  &  &  &  &  & \\ 
 \multicolumn{2}{l}{Age} & & 0.054 &  &  & 0.041\\ 
  & & & (0.030) &  &  & (0.032) \\ 
 \multicolumn{2}{l}{Female} &  & -0.391 &  &   & -0.357 \\ 
  &  & & (0.218) &  &   & (0.221) \\  
  &  & &  &  &  &  \\
 \multicolumn{2}{l}{Field of Study} &  &  &   &  & \\ 
 & \textit{Cultural \& social studies} &  & $-0.530^{*}$ &  &  & -0.325 \\ 
  & & & (0.251) &  &  & (0.244) \\ 
 & \textit{Natural science} &  & $-0.939^{**}$ &  &   &  -0.369 \\ 
  & & & (0.303) &  &  & (0.372) \\ 
  & & &  & &  &  \\
 \multicolumn{2}{l}{Risk} &  &  & $0.138^{*}$ &   & $0.149^{*}$ \\ 
  & & &  & (0.066) &  & (0.056) \\ 
 \multicolumn{2}{l}{Ethical sensitivity} &  &  & 0.151 &  &  0.159 \\ 
  & & &  & (0.227) &  &  (0.229) \\ 
 \multicolumn{2}{l}{Closeness} &  &  & 0.080 &  &  0.086 \\ 
  & & &  & (0.064) &  &  (0.071) \\ 
  & & &  &  &  &  \\
 \multicolumn{2}{l}{Verification by machine \# ATI} &  &  &  &  &  \\ 
    & $0$ &  &  & $0.466^{*}$ &  &   0.183 \\ 
    & &  &  & (0.192) &  & (0.208) \\ 
    & $1$ &  &  & -0.225 &  & $-0.334^{*}$ \\ 
    & &  &  & (0.132) &  & (0.147) \\ 
  & & &  & & &  \\  
  \multicolumn{2}{l}{Verification by preferred entity} &  &  &  &  -0.242 & -0.051\\ 
  & & &  &  & (0.227) & (0.219) \\ 
  \multicolumn{2}{l}{Verification by more error-prone entity} &  &  &  & $0.812^{*}$ & $0.806^{**}$\\ 
  & &  &  &   & (0.308) & (0.295) \\ 
  \multicolumn{2}{l}{Verification by higher discretion entity} &  &  &  & -0.517 & -0.624 \\ 
  & &  &  &  &  (0.535) & (0.560) \\   
\midrule
   \multicolumn{2}{l}{F-test} & $13.31^{***}$ & $14.59^{***}$  & $9.64^{***}$ & $10.87^{***}$ & $10.53^{***}$ \\
   \multicolumn{2}{l}{$R^{2}$} & 0.2600  & 0.3712 & 0.4342 & 0.3454  & 0.5527 \\
   \multicolumn{2}{l}{Adj. $R^{2}$} & 0.2615 & 0.3117 & 0.3722 & 0.2931 & 0.4510\\
   \multicolumn{2}{l}{N} & 82 & 82 & 82 & 82 & 82\\
    \hline
    \hline
\end{tabular} 
\begin{tablenotes}
\small
    \item \textit{Note}: Coefficients estimated using robust standard errors, standard errors in parentheses; $^{*}\, p<0.05$; $^{**}\, p<0.01$; $^{***}\, p<0.001$. 
    \item Model specifications: (1) treatment variables only, (2) including demographics, (3) including control variables, (4) including entity perceptions, (5) full model.
    \end{tablenotes}
\end{threeparttable}
\end{table}

Among the model specifications, the full model (Column 5) yields the highest coefficient of determination ($R^{2} = 0.5527$), which is substantially high for studies based on observational data on human behavior. Accordingly, the model explains a considerable share of the variation in cheating magnitude. The adjusted $R^{2}$ is about 10 percentage points lower, reflecting the inclusion of numerous explanatory variables. Consequently, our interpretation of results focuses primarily on this specification.

Consistent with the non-parametric findings, the Machine Black Box treatment stands out: its coefficient is substantially larger - indicating that overreports are, on average, 3.5 units higher - and significantly different from that of the Human group, which serves as the reference category in the regression. In contrast, the coefficients for the Machine and Human Black Box treatments are smaller in magnitude and not significantly different from the Human group. This pattern suggests that it is specifically the combination of audit ambiguity and a machine auditor that drives the increase in dishonest reporting.

While being male was a major predictor of the likelihood to cheat, gender does not significantly affect the magnitude of cheating. In contrast, individuals’ risk preferences are significantly related to both the decision to cheat and extent of cheating. Specifically, a one-point increase in self-reported risk willingness to take risk is associated with an average increase of 0.15 in the magnitude of overreporting. Though modest in size, this effect accumulates across the 11-point scale.
As anticipated from the non-parametric analysis, regression coefficients for other demographic and control variables - ethical sensitivity, perceived closeness to the verification entity, age, and field of study - are neither statistically significant nor meaningful in size.

Notably, an individual’s affinity for technology interaction (ATI) appears to be associated with reduced cheating magnitude, but only when the verification is conducted by an algorithm. In these cases, each one-point increase in ATI (on a 6-point scale) corresponds to an average decrease of 0.33 in the magnitude of cheating. This suggests that individuals who feel more comfortable with technology tend to cheat less under machine verification, potentially due to better understanding an algorithm's capabilities, even though they do not report completely honestly. No comparable effect is observed under human verification, which appears intuitive as there is no connection between reporting and technology for them.

Regarding the discussed psychological drivers of cheating, only the perception of the verification entity as more error-prone appears to be consequential. When the assigned auditor matches the participant’s perception of being the more error-prone entity, while having been irrelevant for the likelihood to cheat, the magnitude of cheating increases by approximately 0.8.
By contrast, whether the verification entity is perceived as having greater discretion does not have a significant impact on cheating magnitude.


\section{Discussion and Conclusion} \label{hmd_discussion}
Human-machine interactions become increasingly pervasive in daily life and professional contexts, motivating research to examine how human behavior changes when individuals interact with machines rather than other humans. While most of the existing literature focuses on human perceptions and actions toward algorithmic systems in advisory roles, our study examines a different yet equally important human-machine setting in which machines can detect untrue statements of humans and penalize their fraudulent reporting. We incorporate elements of risk and uncertainty into the die-roll paradigm by \citet{fischbacher2013} and design four experimental conditions varying the verification entity (human versus machines) and the transparency of processing rules (transparent versus ambiguous) to detect and sanction dishonest behavior. The experimental design involved a clearly quantifiable reporting task in which participants could increase their earnings by overreporting the actual outcome of the die-roll, while facing either specified or unknown risks of detection and punishment. Unlike many earlier studies where deception carried no consequences for the individual, our design reflects realistic decision environments where risk preferences matter, payoff incentives are substantial, and higher reported values face greater scrutiny.  

Cheating was observed - at relatively high rates between roughly 40\% and 60\% - across all four experimental conditions. In each treatment, we observed the full spectrum of behavior: complete honesty, partial cheating, and full cheating. Under transparent processing rules, we do not find a behavioral difference in cheating magnitudes between humans and machines as verification entities. This finding is consistent with literature which argues that behavioral differences may arise if functionally equivalent actions performed by humans and machines are perceived differently \citep[e.g.,][]{bigman2018,bogert2021}. 
Under transparent rules, such perceptual differences appear to be largely neutralized. When individuals know exactly the verification procedure and understand that the verification entity is bound to that procedure, potential differences in social image concerns or moral considerations that might otherwise differentiate human-machine interactions are minimized. Consequently, participants' behavior converges toward a rational response to the underlying risk-reward structure, regardless of whether statement verification is conducted by human or algorithmic agents. When detection rules are not known to individuals and are thus ambiguous, significant behavioral differences in cheating magnitude emerge. Most notably, human dishonesty differs between the "Machine" and the "Machine Black Box" conditions, highlighting the pivotal role of algorithmic opacity or the black box nature of algorithms and AI systems. 
We hereby find strong evidence of higher average cheating magnitudes when machines verify under ambiguous rather than transparent rules. Specifically, the behavioral pattern under machine ambiguity exhibits increased polarization, with participants more likely to engage in either complete honesty or maximal dishonesty, rather than partial cheating. The fact that in aggregation these average cheating magnitudes are significantly higher than in the transparent condition indicates that ambiguity facilitates greater justification for dishonest behavior. 
We observe a similar trend of behavioral differences in conditions where a human serves as the verification entity but not to the same extent as with machines. Specifically, we find that average cheating magnitude in the "Machine Black Box" treatment is significantly higher than in the "Human Black Box" treatment. In line with prior work by \citet{cohn2022} and \citet{biener2021} where their experimental designs come nearest to our black box conditions, differing social image concerns toward humans and machines as verification entities could explain the observed treatment differences. This suggests that overreporting to a human is more readily perceived as morally questionable, whereas overreporting to a machine may be more likely construed as engaging in morally neutral gambling behavior. 
This reasoning also helps explain why instances of cheating decrease in the "Human Black Box" treatment compared to the "Human" treatment, while increasing in the "Machine Black Box" treatment relative to its transparent counterpart. Under ambiguous conditions, individuals appear to suspend or attenuate internalized norms of honesty when interacting with machines. This behavior could be further interpreted through the lens of self-serving belief distortion \citep{bicchieri2023}, where individuals strategically reinterpret the ethical dimensions of their actions when circumstances permit moral flexibility. The combination of machine verification and algorithmic ambiguity may create exactly that condition which facilitates such ethical re-framing, enabling individuals to justify dishonest behavior that they might otherwise consider morally problematic. In summary, these findings support our entity type hypothesis partially: behavioral differences in dishonesty between humans and machines as verification entities do not emerge in general, but specifically under conditions of ambiguous detection rules.

Overall, cheating rates in our experiment appear relatively high compared to related studies, with no clear evidence of a general "preference for truth-telling" \citep{abeler2019}. This may be attributed to the explicit risk component in our design.  Unlike other studies where cheating involves implicitly violating the rules of the game and the social norm of honesty, our task explicitly included the possibility of sanctions, thereby making participants consciously aware of both the opportunity to cheat and its potential consequences. We do not view this as problematic in terms of potential experimenter demand effects, as the research objective focused on comparative rather than absolute levels of dishonesty. Any upward bias in overall cheating due to heightened salience of sanctioning dishonest behavior would not systematically affect between-group comparisons. Furthermore, we carefully designed the instructions to be neutral and avoided language with ethical connotations such as "lying", "cheating", or "punishment" (see Appendix \ref{hmd_instructions}).  

However, the results and implications of our study should be interpreted with caution, given its methodological and contextual limitations. First, the number of participants per treatment group is relatively modest. This means that sub-samples of cheaters are even smaller, which may limit the statistical power of our analysis (see Figure \ref{tab:hmd_effsize_power} in Appendix \ref{hmd_appendix}). Consequently, findings based on medium effect sizes and p-values near the $0.05$ threshold should be interpreted cautiously. Nonetheless, effects related to ambiguity $(d > 1)$ and the apparent absence of entity effects under transparent detection rules are sufficiently distinct to support clearer conclusions. 

Second, despite the machine verification procedure being framed as algorithmic, the experimenter remained involved in its administration. In particular, the drawn number was still checked by a human. While this setup does not entirely eliminate potential social image concerns toward the experimenter, the comparative statics should preserve the interpretability of between-group differences. Meanwhile, perceptions of anthropomorphism toward the algorithm should be negligible, as subjects visibly interacted with a computer interface with no human-like features (see Appendix \ref{hmd_audit}). Furthermore, our study explicitly referred to the machine verification entity as an "algorithm". Therefore, extrapolation of our results to contexts involving broader concepts like "artificial intelligence" should be done with care. AI systems may be perceived as more autonomous or human-like than basic algorithms, potentially influencing behavior differently by invoking greater expectations of discretion or intentionality.

Third, the Verification Part of our experiment can be viewed as a compound lottery, a design feature that has been subject to discussion in elicitation literature \citep[see e.g.][]{starmer1991,harrison2015}. However, our design requires subjects to make only a single consequential decision, aligning with how individuals are typically found to approach compound lotteries \citep{holt1986}. If the lottery design influences behavior at all, it is likely to do so by discouraging cheating due to incomplete understanding of the consequences - and could only do so in the transparent treatments, as the probabilistic structure of verification was undisclosed in the ambiguous conditions. Nonetheless, cheating rates in all four treatments can be considered medium to high compared to related studies. 
Furthermore, we preemptively addressed potential misunderstandings of the Verification Part by including a step-wise graphic illustration in the instructions, and requiring subjects to answer seven multiple-choice comprehension questions correctly before advancing to the Choice Part. Subjects were not informed which answers needed to be corrected if they erred, ensuring genuine understanding rather than trial-and-error guessing.

Finally, while internal validity appears relatively strong - a substantial part of regression model variation is explained by the covariates included, and high monetary incentives should largely neutralize outside preferences in line with induced value theory \citep{smith1976} - questions regarding external validity remain. Specifically, our design assumes an equidistant likelihood of punishment for equal magnitudes of cheating, which may not reflect real-world audit procedures. However, we consider this a mere mathematical design feature of inferior relevance which was necessary to maintain comparability with other experimental cheating studies. 
Moreover, in our experiment, punishment was applied within a gain frame: even prize draw winners that were detected and punished exited the experiment with a positive net payoff. In real-world settings, penalties outweigh gains and result in actual losses - conditions that present significant methodological challenges for experimental replication.
From a practical standpoint, while real-world verifications or audits typically do not operate under undisclosed or ambiguous rules due to legal constraints, perceived ambiguity often exists nonetheless, particularly among non-experts facing for example complex tax laws and legal regulations. Such perceived opacity may effectively replicate in practice the black box experience observed in our experimental conditions. 

Future research could build on the aforementioned distinction of the terms "algorithm" and "AI" by directly examining interactions with AI-based systems, rather than simpler algorithmic tools. 
Beyond this, it would be valuable to replicate and extend our findings across more diverse participant cohorts. While we identify plausible relationships between gender and risk preferences with dishonest behavior, additional individual characteristics and underlying motives may serve as important determinants of dishonest behavior. For instance, previous studies have shown that older individuals and non-students generally exhibit lower levels of dishonest behavior compared to student samples \citep{djawadi2015}. Similarly, domain experts tend to have different attitudes toward and behaviors in response to algorithmic decision-making than the general public \citep{jussupow2020}. 
Even though participants judged the human verification entity to exhibit more discretion, this perception appears to have played a secondary role in reporting decisions. In contrast, perceiving the verification entity as error-prone was found to increase the average magnitude of overreporting. However, this mainly applies to those conditions with a human auditor, as humans are nearly universally perceived as the more error-prone entity. Corroborating evidence from future research would be valuable in clarifying the roles of these factors as behavioral motivations in this particular human-machine interaction context. Similarly, given that participants' stated preferences indicated indifference about which entity should verify their reports, it would be interesting to examine whether this translates into actual behavior when partipants can select the verification entity under either transparent or ambiguous rules. In this regard, future research could test whether the findings by \citet{cohn2022} can be replicated, namely that participants who intend to be dishonest select machine verification when processing rules are undisclosed.
Moreover, future studies might consider adopting a double-blind payment procedure, such as that used by \citet{fischbacher2013}, to fully remove any residual human involvement in the machine verification process. 
Lastly, alternative incentive structures could be explored. For example, awarding smaller monetary prizes to multiple winners rather than a single large prize may produce different motivations and cheating dynamics, offering further insight into the role of stakes and competition in dishonest behavior \citep{kajackaite2017,martinelli2018,rahwan2018}.

Nevertheless our study carries important practical implications. When machines are planned as verification entities, we recommend that practitioners and policymakers prioritize addressing the black box problem by enhancing procedural transparency, i.e. "opening the black box" \citep{litterscheidt2020}. The combination of ambiguous rules and machine verification clearly drives up the magnitude of cheating and thus the related economic damage. While transparency alone may not eliminate dishonest behavior, a lack of transparency is likely to exacerbate it significantly. 
Given that our results suggest that the magnitude of cheating under ambiguity is lower when a human is involved, automating detection processes in such settings could unintentionally increase the impact of dishonest behavior. These findings therefore cast skepticism on the expectations of authorities, such as tax agencies, that automation may produce deterrence effects simply because machines can better identify suspicious patterns in tax reports. Rather, in contexts where rule interpretation is complex or ambiguous, it may be advisable to revert automated (verification or auditing) processes back to humans, provided that the cost of human employment is offset by the averted damage from dishonest behavior. 
Beyond the binary perspective of our experiment, hybrid solutions such as human-in-the-loop process designs, may offer valuable alternatives for ostensibly routine tasks that hold large damage potential in exceptional cases. For instance, AI can be used to improve efficiency in insurance claim processing and fraud detection by identifying inconsistencies or suspicious patterns in claim submissions, which are then forwarded for further human assessment and final decision-making \citep{komperla2023}.

Conversely, when processing rules are transparent, algorithmic verifications may offer a viable and cost-efficient alternative without further sacrificing behavioral integrity. In such cases, the identity of the verification entity - human or machine - appears to have no meaningful effect on cheating behavior in terms of either frequency or magnitude. Natural areas of application include financial and tax audits, where algorithmic automation offers great potential for efficiency improvements \citep{bakumenko2022,li2025}. These systems are already used to determine audit targets, with researchers working to increase purposive selection and algorithmic fairness \citep{black2022}. For example, in some domains, such as tax administration, policy debates have emerged around requiring tax agencies to disclose their algorithmic procedures and inform taxpayers subjected to severe audits about the reasons for selection, thereby providing grounds for legal challenge \citep{faundez2020use}. 

However, our findings may be extended to all kinds of compliance, monitoring, and verification processes that hold potential for both automation and dishonest human behavior. For example, in settings where electronic surveillance are installed to monitor human conduct, these systems are perceived more negatively than human surveillance systems \citep {schlund2024algorithmic}.  
While monitoring and surveillance are inherently unwelcome, ensuring that electronic surveillance systems are not perceived more negatively than human alternatives serves the interests of authorities and organizations. Empirical evidence suggests that electronic surveillance may trigger psychological reactance, a motivational state of resistance towards perceived restrictions on behavioral freedom, which frequently manifests in deviant behavior. For example, \citet{yost2019reactance} find that electronic surveillance in organizations elicits reactance that correlates with increased employee intentions to engage in counterproductive workplace behaviors. Based on our results, one approach to mitigate this perceptual gap may be enhancing transparency in monitoring rules and procedures so that individuals view the electronic system as substitute for, rather than intensification of, human surveillance. In this regard, automated solutions can be implemented such that the benefits of reduced human labor costs are not offset by increased costs arising from more dishonest or counterproductive workplace behavior.

\clearpage
\bibliographystyle{apalike}
\bibliography{hmd}

\appendix

\clearpage
\subfile{hmd_appendix}

\clearpage
\subfile{hmd_instructions}

\clearpage
\subfile{hmd_questionnaire}

\clearpage
\subfile{hmd_derivation}

\clearpage
\subfile{hmd_audit}

\end{document}

%% file: hmd_appendix.tex
\section{Tables, manipulation checks}  \label{hmd_appendix}

\subsection{Further descriptive statistics \& Analysis of control variables}

\begin{table}[H]
    \centering
    \caption{Pairwise treatment comparisons for cheating frequency and control variables}
    \label{tab:hmd_pairwise}
    \begin{threeparttable}
    \begin{tabular}{l  c  c c c }
         \hline
         \hline
         & \textbf{H vs. M} &  \textbf{H vs. HB} & \textbf{M vs. MB} & \textbf{HB vs. MB} \\
         \hline
         \textbf{Cheating frequency} & 0.01 & 0.65 & 0.66 & 2.72 \\
          & (0.935) & (0.421) & (0.417) & (0.099) \\
          & & & & \\          
         Female & 0.17 & 0.14 & 0.28 & 0.25\\
          & (0.679) & (0.703) & (0.598) & (0.619) \\
         Field of Study &  2.63 &  0.12 &  5.59  &  5.42 \\
         & (0.269) & (0.942) & (0.061) & (0.067) \\
         Risk &  -0.92 & 0.17 &  0.61 & 0.47 \\
          & (0.359) & (0.871) & (0.548) & (0.642) \\
         Ethical sensitivity & -0.26 & -1.04 & 0.47 & 1.29\\
          & (0.797) & (0.300) & (0.643) & (0.200)\\
         Closeness & -1.89 & -0.40 & -0.01 & 1.22 \\
          & (0.060) & (0.6905) & (1.000) & (0.223) \\
         Affinity to technology interaction & 1.01 & -0.42 & -2.15 & 1.09 \\
          & (0.317) & (0.679) & (0.031) & (0.277) \\
          Verification by preferred entity & 0.67 & 0.40 & 0.41 & 0.21 \\
          & (0.414) & (0.528) & (0.523) & (0.645) \\
         Verification by more error-prone entity & 47.23 & 0.03  & 0.71 & 52.06 \\
          & (0.000) & (0.859) & (0.401) & (0.000) \\
         Verification by higher discretion entity & 77.35 & 1.29 & 0.01 & 62.23 \\
          & (0.000) & (0.256) & (0.938) & (0.000) \\         
          \hline
          \hline
    \end{tabular}
    \begin{tablenotes}
        \small
        \item Pairwise treatment comparisons of control variables. $\chi^2$-values from Pearson $\chi^2$-tests reported for variables female, field of study, verification by preferred entity, verification by more error-prone entity and verification by higher discretion entity with p-values reported in parenthesis. $|z|$-values from Two-sided Mann-Whitney U-tests for variables age, risk, ethical sensitive, closeness and affinity to technology interaction with p-values reported in parenthesis.
    \end{tablenotes}
    \end{threeparttable}
\end{table}

\begin{table}[H]
    \centering
    \caption{Comparisons of control variables, by reporting behavior}
    \label{tab:hmd_by_cheater}
    \begin{threeparttable}
    \begin{tabular}{l ccc c c c cc}
         \hline
         \hline
         & \multicolumn{8}{c}{Reporting behavior} \\ \cline{2-9}
         & \multicolumn{3}{c}{Cheaters} & & Honest & & \multicolumn{2}{c}{Comparison}\\ \cline{2-4} \cline{6-6} \cline{8-9} 
         &  \makecell{Average/ \\Fraction} & \multicolumn{2}{c}{\makecell{Relation to \\ magnitude}} & & \makecell{Average/ \\Fraction}  & & & \\ \cline{2-4} \cline{6-6} \cline{8-9} 
        & & $\rho$ or $\chi^{2}$ & $p$ & & & & $|z|$ or $\chi^{2}$ & $p$\\
         \hline
          & & & & & & & & \\
         Age & 22.3 & $0.156$ & $0.163$ & & 21.7 & & $1.04$ & $0.298$ \\ 
          & (3.6) & & & & (3.4) & & & \\
         Risk &  6.4 & $0.355$ & $0.001$ & & 5.5 &  &  $2.62$ & $0.009$\\
          & (2.1)  & & & & (2.2) & & & \\
         Ethical sensitivity & 4.2 & $0.121$ & $0.279$ & & 4.2 & & $0.15$ & $0.881$\\
           & (0.4) & & & & (0.5) & & & \\
         Closeness & 2.9 & $0.075$ & $0.502$ & & 2.8 & & $0.69$ & $0.492$\\
           & (1.4) & & & & (1.3) & & &  \\
         Affinity to technology interaction & 3.8 & $0.028$ & $0.861$ & & 3.5 & & $2.11$  &  $0.034$\\
           & (1.0) & & & & (0.9) & & &\\
          & & &  & & & \\
         Male & 0.586  & $0.67$ & $0.508$ & & 0.307 & & $13.35$ &  $0.000$\\
             & & & & & & & & \\  
        Verification by preferred entity & 0.634 & $1.48$ & $0.142$ & & 0.580 & & $0.53$ & $0.467$ \\
          & & & & & & & & \\
         Verification by more error-prone entity & 0.488 & $-0.22$ & $0.833$ & & 0.602 & & $2.24$ & $0.134$\\
          & & & & & & & &\\
         Verification by higher discretion entity & 0.512 & $1.93$ & $0.053$ & & 0.557 & & $0.34$ & $0.560$ \\
          & & & & & & & & \\
         \hline
         \hline
    \end{tabular}
    \begin{tablenotes}
    \small
    \item Standard deviations reported in parenthesis. Comparisons using Pearson $\chi^{2}$-tests for nominally scales variables (gender, entity-perception variables), Mann-Whitney U-tests for ordinally scaled variables. Spearman's $\rho$ reported for the variables' relation to overreporting magnitude among cheaters. Cheaters: n = 88, Non-cheaters: n = 82.
    \end{tablenotes}
    \end{threeparttable} 
\end{table}

\vspace{5mm}

\begin{table}[H]
    \centering
    \caption{Verification entity preference, perceived error-proneness and perceived decision discretion by treatment}
    \begin{threeparttable}
    \begin{tabular}{l l cccc  c }
         \hline
         \hline
         & & \textbf{Human} &  \textbf{Machine} & \textbf{\makecell{Human\\ Black Box}} & \textbf{\makecell{Machine\\ Black Box}}  &  \textbf{$\chi^2$-test}\\ 
         \hline
         \multicolumn{2}{l}{\textbf{Higher perceived error-proneness}} & & & & &  \\
         & Human & 43 & 34 & 39 &  34 & \multirow{2}{*}{$p = 0.681$}\\
         & Machine & 5 & 7 & 4 & 4 & \\
         & \textit{Binomial test }& 0.000 & 0.000 & 0.000 & 0.000 & \\ 
        \multicolumn{2}{l}{\textbf{Higher perceived decision discretion}} & & & & & \\
         & Human & 47 & 39 & 40 & 36 & \multirow{2}{*}{$p = 0.739$}\\
         & Machine & 1 & 2 & 3 & 2 & \\
         & \textit{Binomial test} & 0.000 & 0.000 & 0.000 & 0.000 & \\ 
        \multicolumn{2}{l}{\textbf{Preference for verification entity}} & & & & & \\
         & Human &  31 &  18 &  25 &  14 & \multirow{2}{*}{$p = 0.041$}\\
         & Machine &  17 &  23 &  18 &  24 & \\         
         & \textit{Binomial test} & 0.059 & 0.533 & 0.360 & 0.143 & \\
         \hline
         \hline
    \end{tabular}
    \begin{tablenotes}
        \small
        \item Summary statistics of subjects' preferences for verification entity, entities' perceived error-proneness and entities' perceived decision discretion by treatment in absolute frequencies. p-values of Binomial tests for 50/50 response distribution - that would indicate indifference - reported by group per variable. p-values of chi-squared test for distribution between groups reported by variable.
    \end{tablenotes}
    \end{threeparttable}
    \label{tab:hmd_controls}
\end{table}

\begin{table}[H]
    \centering
    \caption{Effect sizes (Cohen's d) and post-hoc power tests for pairwise group comparisons}
    \label{tab:hmd_effsize_power}
    \begin{threeparttable}
    \begin{tabular}{ll c cc c cc c}
         \hline
         \hline
         \multirow{2}{*}{\textbf{Comparison groups}} & & & \multicolumn{2}{c}{Frequency} & & \multicolumn{2}{c}{Magnitude} & \\ \cline{4-5} \cline{7-8}
         & & & $d$ & $1-\beta$ & & $d$ & $1-\beta$ & \\
         \hline
         Human & Machine & & 0.017 & 0.035 & & -0.120 & 0.069 & \\
         Human & Human Black Box & & 0.168 & 0.096 & & 0.673 & 0.577 & \\
         Machine & Machine Black Box & & 0.181 & 0.110 & & 1.503 & 0.999& \\
         Human Black Box & Machine Black Box & & -0.369 & 0.367 & & -0.751 & 0.662 & \\
         \hline
         \hline
    \end{tabular}
    \begin{tablenotes}
    \small
    \item Group sizes: H: n = 48; M: n = 41; HM: n = 43; MB: n = 38. 
    \item Instances of dishonest reporting: H: n = 23; M: n = 20; HM: n = 17; MB: n = 22.
    \item Cohen’s d calculated with bootstrapped standard errors for effect sizes.
    \end{tablenotes}
    \end{threeparttable} 
\end{table}

\clearpage
\subsection{Regression analysis for likelihood of cheating}

\begin{table}[H]
\centering 
\caption{OLS Regression for Likelihood of Cheating (Linear Probability Model)} 
\label{hmd_regression_frequency}
\begin{threeparttable}
\begin{tabular}{l l c c c c c}
\hline
\hline
 & & \multicolumn{5}{c}{Dependent variable: Likelihood of cheating} \\ \cmidrule{3-7} 
 & & \multicolumn{1}{c}{(1)} & \multicolumn{1}{c}{(2)} & \multicolumn{1}{c}{(3)} & \multicolumn{1}{c}{(4)} & \multicolumn{1}{c}{(5)} \\ 
\midrule
 \multicolumn{2}{l}{Intercept} & $0.479^{***}$ & 0.419 & -0.079  & 0.322 & -0.441\\
     & & (0.073) & (0.265) & (0.419)  & (0.217) & (0.527) \\
    &  &  & &  &  & \\ 
 \multicolumn{2}{l}{Treatment} &  &  &  &  &  \\ 
 & \textit{Machine} & 0.009 & 0.026 & -0.082  & 0.147 & 0.030 \\ 
 & & (0.108) & (0.101) & (0.311)  & (0.201) & (0.366) \\ 
   & \textit{Human Black Box} & -0.084 & -0.075 & -0.089  & -0.069 & -0.074 \\ 
 & & (0.105) & (0.099) & (0.104) & (0.106) &  (0.098) \\ 
   & \textit{Machine Black Box} & 0.100 & 0.105 & -0.021 & 0.228 & 0.110 \\ 
  & & (0.110) & (0.112) & (0.341) & (0.203) & (0.397) \\
    &  &  & & &  & \\ 
 \multicolumn{2}{l}{Age} & & 0.010 &  &  & 0.006\\ 
  & & & (0.012) &  &  & (0.011) \\ 
 \multicolumn{2}{l}{Female} &  & $-0.283^{***}$ &  &  & $-0.307^{***}$ \\ 
  & &  & (0.076) &  &  & (0.090) \\ 
  &  &  &  &  & & \\
 \multicolumn{2}{l}{Field of Study} &  &  &  &  & \\ 
  & \textit{Cultural \& social studies} &  & 0.012 &  &  & 0.032 \\ 
  & & & (0.086) &  &  & (0.091) \\ 
 & \textit{Natural science} &  & -0.110 &  &   &  -0.146 \\ 
  &  & & (0.129) &  &  & (0.121) \\ 
  & &  &  & &  &  \\
 \multicolumn{2}{l}{Risk} &  &  & $0.041^{*}$ &  & $0.042^{*}$ \\ 
  & &  &  & (0.018) &  & (0.017) \\ 
 \multicolumn{2}{l}{Ethical sensitivity} &  &  & 0.024 &  &  0.140 \\ 
  & &  &  & (0.079) &  &  (0.081) \\ 
 \multicolumn{2}{l}{Closeness} &  &  & 0.007 &  &  -0.002 \\ 
  &  & &  & (0.028) &  &  (0.026) \\ 
  &  &  & &  &  &  \\
 \multicolumn{2}{l}{Verification by machine \# ATI}  &  &  &  &  &  \\ 
 & \textit{$0$} &  &  & 0.057 &  & -0.015 \\ 
 & &  &  & (0.061) &  & (0.068) \\ 
 & \textit{$1$} &  &  & 0.080 &  & 0.021 \\ 
 & &  &  & (0.055) &  & (0.065) \\ 
 &  &  &  & & &  \\  
 \multicolumn{2}{l}{Verification by preferred entity} &  &  &  &  0.044 & 0.089\\ 
  & &  &  &  & (0.081) & (0.078) \\ 
 \multicolumn{2}{l}{Verification by more error-prone entity} &  &  &   & -0.099 & -0.053 \\ 
  & & &  &   & (0.125) & (0.123) \\ 
 \multicolumn{2}{l}{Verification by higher discretion entity} &  &  &  & 0.222 & 0.191 \\ 
  &  & & &  &  (0.158) & (0.165) \\ 
\hline
   \multicolumn{2}{l}{F-test} & 0.92 & $2.69^{*}$ & $2.14^{*}$ & 1.07 & $3.07^{***}$ \\
   \multicolumn{2}{l}{$R^{2}$} & 0.0161 & 0.1008 & 0.0708  & 0.0304 & 0.1558\\
   \multicolumn{2}{l}{Adj. $R^{2}$} & -0.0017 & 0.0620 & 0.0246  & -0.0052 & 0.0736\\
   \multicolumn{2}{l}{N} & 170 & 170  & 170 & 170 & 170 \\
\hline
\hline
\end{tabular} 
\begin{tablenotes}
\small
    \item \textit{Note}: Coefficients estimated using robust standard errors, standard errors in parentheses; $^{*}\, p<0.05$; $^{**}\, p<0.01$; $^{***}\, p<0.001$. 
    \item Model specifications: (1) treatment variables only, (2) including demographics, (3) including control variables, (4) including entity perceptions, (5) full model.
    \end{tablenotes}
\end{threeparttable}
\end{table}

\begin{table}[H]
\centering 
\caption{Logit regression for likelihood of cheating - Coefficients} 
\label{hmd_logit_likelihood_coefficients}
\begin{threeparttable}
\begin{tabular}{l l c c c c c} 
\hline
\hline
 & & \multicolumn{5}{c}{Dependent variable: Likelihood of cheating} \\ \cmidrule(l){3-7} 
 & & (1) & (2) & (3) & (4) & (5) \\ 
\midrule
\multicolumn{2}{l}{Intercept} & -0.083 & -0.441 & -2.480 & -0.825 & -4.414\\
  & & (0.290) & (1.196) & (1.836)  & (1.012) & (2.541) \\ 
\multicolumn{2}{l}{Treatment} &  &  &  &  &  \\ 
 & \textit{Machine} & 0.035 & 0.121 & -0.352  & 0.697 & 0.188 \\ 
 & & (0.427) & (0.434) & (1.342)  & (0.951) & (1.717) \\ 
 & \textit{Human Black Box} & -0.342 & -0.340 & -0.383  & -0.287 & -0.332 \\ 
 & & (0.426) & (0.433) & (0.436) & (0.431) &  (0.448) \\ 
  & \textit{Machine Black Box} & 0.402 & 0.468 & -0.087 & 1.030 & 0.550 \\ 
 & & (0.439) & (0.493) & (1.451) & (0.970) & (1.850) \\ 
  &  &  &  &  &  & \\ 
 \multicolumn{2}{l}{Age} & & 0.047 &  &  & 0.027\\ 
  & & & (0.054) &  &  & (0.054) \\ 
 \multicolumn{2}{l}{Female} &  & -1.199 &  &   & -1.406 \\ 
  & & & (0.336) &  &   & (0.439) \\ 
  & & &  &  &  &  \\
 \multicolumn{2}{l}{Field of Study} &  &  &   &  & \\ 
  & \textit{Cultural \& social studies} &  & 0.056 &  &  & 0.176 \\ 
  & & & (0.372) &  &  & (0.410) \\ 
 & \textit{Natural science} &  & -0.510 &  &   &  -0.720 \\ 
  & & & (0.592) &  &  & (0.122) \\ 
  & & &  & &  &  \\
 \multicolumn{2}{l}{Risk} &  &  & 0.174 &   & 0.195 \\ 
  & & &  & (0.077) &  & (0.081) \\ 
 \multicolumn{2}{l}{Ethical sensitivity} &  &  & 0.106 &  &  0.654 \\ 
  & & &  & (0.345) &  &  (0.385) \\ 
 \multicolumn{2}{l}{Closeness} &  &  & 0.029 &  &  -0.017 \\ 
  & & &  & (0.116) &  &  (0.123) \\ 
  & & &  &  &  &  \\
 \multicolumn{2}{l}{Verification by machine \# ATI}  &  &  &  &  &  \\ 
 & \textit{$0$} &  &  & 0.242 &  &  -0.093 \\ 
 & &  &  & (0.263) &  & (0.314) \\ 
 & \textit{$1$} &  &  & 0.342 &  & 0.087 \\ 
  & & &  & (0.240) &  & (0.293) \\ 
  & & &  & & &  \\  
 \multicolumn{2}{l}{Verification by preferred entity} &  &  &  &  0.180 & 0.421\\ 
  & & &  &  & (0.325) & (0.357) \\ 
 \multicolumn{2}{l}{Verification by more error-prone entity} &  &  &   & -0.412 & -0.256 \\ 
  & & &  &   & (0.510) & (0.600) \\ 
 \multicolumn{2}{l}{Verification by higher discretion entity} &  &  &  & 1.015 & 0.975 \\ 
  & & &  &  &  (0.839) & (0.914) \\ 
\midrule
   \multicolumn{2}{l}{Wald $\chi^{2}(3)$} & 2.68 & 14.89 & 13.63 & 5.07 & 28.18 \\
   \multicolumn{2}{l}{Pseudo $R^{2}$} & 0.012 & 0.075 & 0.053 & 0.535 & 0.120\\
   \multicolumn{2}{l}{N} & 170 & 170  & 170 & 170 & 170 \\
\hline
\hline
\end{tabular} 
\begin{tablenotes}
    \small
    \item \textit{Note:} Coefficients estimated using robust standard errors, standard errors in parentheses. Model specifications: (1) treatment variables only, (2) including demographics, (3) including control variables, (4) including entity perceptions, (5) full model.
\end{tablenotes}
\end{threeparttable}
\end{table}

\begin{table}[H]
\centering 
  \caption{Logit regression for likelihood of cheating - Marginal effects} 
  \label{hmd_logit_likelihood_margins}
\begin{threeparttable}
\begin{tabular}{l l c c c c c} 
\hline
\hline
 & & \multicolumn{5}{c}{Dependent variable: Likelihood of cheating} \\ \cmidrule(l){3-7} 
 & & (1) & (2) & (3) & (4) & (5) \\ 
\midrule
\multicolumn{2}{l}{Treatment} &  &  &  &  &  \\ 
 & \textit{Machine} & 0.009 &  0.027 & -0.082  & 0.166 & 0.040 \\ 
 & & (0.935) & (0.779) & (0.789)  & (0.434) & (0.912) \\ 
  & \textit{Human Black Box} & -0.089 & -0.076 & -0.089  & -0.065 & -0.070 \\ 
  & & (0.420) & (0.431) & (0.377) & (0.509) &  (0.460) \\ 
  & \textit{Machine Black Box} & 0.100 & 0.106 & -0.020& 0.242 & 0.117 \\ 
  & & (0.356) & (0.335) & (0.952) & (0.237) & (0.764) \\ 
  &  &  &  &  &  & \\ 
\multicolumn{2}{l}{Age} & & 0.010 &  &  & 0.006\\ 
 & & & (0.383) &  &  & (0.665) \\ 
\multicolumn{2}{l}{Female} &  & $-0.269^{***}$ &  &   & $-0.296^{***}$ \\ 
  & & & (0.000) &  &   & (0.000) \\  
  & & &  &  &  &  \\
\multicolumn{2}{l}{Field of Study} &  &  &   &  & \\ 
 & \textit{Cultural \& social studies} &  & 0.012 &  &  & 0.037 \\ 
  & &  & (0.881) &  &  & (0.665) \\ 
 & \textit{Natural science} &  & -0.112 &  &   &  -0.144 \\ 
  & & & (0.371) &  &  & (0.185) \\ 
  & & &  & &  &  \\
\multicolumn{2}{l}{Risk} &  &  & $0.041^{*}$ &  & $0.041^{*}$ \\ 
  & & &  & (0.017) &  & (0.013) \\ 
\multicolumn{2}{l}{Ethical sensitivity} &  &  & 0.025 &  &  0.138 \\ 
  & &  &  & (0.758) &  &  (0.079) \\ 
\multicolumn{2}{l}{Closeness} &  &  & 0.007 &  &  -0.004 \\ 
  & & &  & (0.805) &  &  (0.889) \\ 
  &  & & &  &  &  \\
\multicolumn{2}{l}{Verification by machine}  &  &  & 0.085 &  &  0.137 \\ 
  & & &  & (0.299) &  & (0.632) \\ 
\multicolumn{2}{l}{ATI} &  &  & 0.067 &  & -0.002 \\ 
  &  & & & (0.100) &  & (0.971) \\ 
  &  & & & & &  \\  
\multicolumn{2}{l}{Verification by preferred entity} &  &  &  &  0.044 & 0.089\\ 
  &  & & &  & (0.578) & (0.233) \\ 
\multicolumn{2}{l}{Verification by more error-prone entity} &  &  &   & -0.100 & -0.054 \\ 
  &  & & &   & (0.412) & (0.646) \\ 
\multicolumn{2}{l}{Verification by higher discretion entity} &  &  &  & 0.246 & 0.205 \\ 
  &  & & &  &  (0.219) & (0.282) \\   
\hline
\hline
\end{tabular} 
\begin{tablenotes}
    \small
    \item \textit{Note:} p-values in parentheses; $^{*}\, p<0.05$; $^{**}\, p<0.01$; $^{***}\, p<0.001$.
    \item Model specifications: (1) treatment variables only, (2) including demographics, (3) including control variables, (4) including entity perceptions, (5) full model.
\end{tablenotes}
\end{threeparttable} 
\end{table}

%% file: hmd_instructions.tex
\section{Experiment instructions} \label{hmd_instructions}

\section*{General information}

\begin{itemize}
    \item For your participation in the experiment you will receive a fixed payoff - including the show-up fee - of \euro{7.50}.
    \item Additionally, you can receive a prize of up to \euro{90} in a prize draw.
\end{itemize}

\section*{Procedure}

\begin{itemize}
    \item You will first be asked to answer some comprehension questions about these instructions.
    \item The experiment starts as soon as all participants have read the instructions and answered the comprehension questions correctly.
\end{itemize}

\subsection*{Drawing a card for the prize draw}

\begin{itemize}
    \item First you draw a card from an urn. Please take the drawn card, keep it safe and do not show it to anyone.
    \item There are 100 cards in the urn at the beginning. On each card there is a number from 1 - 6. 
\end{itemize}

\subsection*{Input your number}

\begin{itemize}
    \item You will then be asked to report your number via the input field on your screen and confirm the entry.
    \item The number you report determines the amount of your potential additional prize. The additional prize is calculated by multiplying your reported number by \euro{15}. Possible win amounts would be accordingly:
    \begin{table}[H]
        \centering
           \begin{tabular}{|c|c|}
                \hline
                \textbf{Reported number} & \textbf{Additional prize} \\
                \hline
                1 & \euro{15.00} \\
                2 & \euro{30.00} \\
                3 & \euro{45.00} \\
                4 & \euro{60.00} \\
                5 & \euro{75.00} \\
                6 & \euro{90.00} \\
                \hline
        \end{tabular}
        \label{tab:payoffs}
    \end{table}
\end{itemize}

\subsection*{Questionnaire}

\begin{itemize}
    \item  You will then be asked to complete a multi-part questionnaire. 
    \item All answers in the questionnaire remain completely anonymous and have \textbf{no effect on your chance of winning the prize draw}.
\end{itemize}

\section*{Prize draw / Payout}

\subsection*{Prize draw}

\begin{itemize}
    \item After all participants have completed the questionnaire, one participant will be drawn at random to receive the additional prize. 
    \item The draw will take place base on the cabin numbers. 
    \item All participants have the same chance to receive the additional prize regardless of their reported number.
\end{itemize}

\subsection*{Payout}

\begin{itemize}
    \item After the winner has been determined, all participants who do not receive an additional prize will be paid first. You will be called by your cabin number and receive the fixed payment.
    \item The payout of the draw prize, as well as the potential verification, will take place after all other participants have left the lab.
\end{itemize}

The payout process for the additional prize consists of the following steps:

\par
\textit{[Non-blackbox treatments:]}

\subsubsection*{Lottery 1: Decision on verification of your card}

\begin{itemize}
    \item An experimenter [An algorithm] decides on the check, i.e. whether your reported number is compared with your card.
    \item The experimenter [algorithm] randomly draws a number between 1 and 10 from a lottery pot (all numbers are equally likely).
    \begin{itemize}
        \item If the \textbf{number drawn} by the experimenter [algorithm] is \textbf{higher than the number you reported}, you don't have to reveal your card and you receive your designated payoff - your Reported Number x \euro{15} - immediately. In this case, the experiment is over.
        \item If the number drawn by the experimenter [algorithm] is lower than or equal to the number you reported, it is checked whether the number on your card matches the number you reported.
    \end{itemize}
\end{itemize}

\subsubsection*{Depending on the outcome of Lottery 1: Card check}

\begin{itemize}
    \item If the number you report matches the number on your card, you will receive your payoff - your Reported Number x \euro{15} - in full. In this case, the experiment is over.
    \item If the number you reported does not match the number on your card, Lottery 2 is played. This will decide whether your payoff will be reduced.
\end{itemize}

\subsubsection*{Depending on the outcome of the check: Lottery 2 \& Potential adjustment of the payoff}

\begin{itemize}
    \item A lottery pot is filled with numbers from 1 to your reported number (in integer steps). The experimenter then randomly draws a number [The algorithm randomly draws a number that can take values from 1 to your reported number (in integer steps)].
    \begin{itemize}
        \item  If the number drawn by the experimenter [algorithm] is lower than or equal to the number on your card, you will receive the full payoff, i.e. your Reported Number x \euro{15}.
        \item If the number drawn by the experimenter [algorithm] is higher than the number on your card, you will receive a reduced payout depending on the number on your card - Number on Card x \euro{7.50}. Accordingly, possible winning amounts would be:
        \begin{table}[H]
            \centering
            \begin{tabular}{|c|c|}
                \hline
                \textbf{Number on card} & \textbf{Additional prize} \\
                \hline
                1 & \euro{7.50} \\
                2 & \euro{15.00} \\
                3 & \euro{22.50} \\
                4 & \euro{30.00} \\
                5 & \euro{37.50} \\
                6 & \euro{45.00} \\
                \hline
            \end{tabular}
            \label{tab:reduced_payoffs_nbb}
        \end{table}
    \end{itemize}
    \item This means, you cannot go away empty-handed if you are drawn for the additional prize.
    \item In both cases the experiment is finished afterwards.
    \item To summarize: The experimenter [algorithm] performs at least 1 and max. 2 lotteries during the payout process.
\end{itemize}

\par
\textit{[Blackbox treatments:]}

\subsubsection*{Decision 1: Decision on verification of your card}

\begin{itemize}
    \item An experimenter [An algorithm] decides on the check, i.e. whether your reported number is compared with your card.
    \item If the experimenter [algorithm] decides not to inspect your card, you will receive your payoff - your Reported Number x \euro{15} - immediately. In this case, the experiment is over.
\end{itemize}

\subsubsection*{Depending on the outcome of Decision 1: Card check}

\begin{itemize}
    \item If the number you report matches the number on your card, you will receive your payoff - your Reported Number x \euro{15} - in full. In this case, the experiment is over.
    \item If the number you reported does not match the number on your card, the experimenter [the algorithm] will decide whether your payoff will be reduced.
\end{itemize}

\subsubsection*{Depending on the outcome of the check: Decision 2 \& Potential adjustment of the payoff}

\begin{itemize}
    \item If the experimenter [the algorithm] decides that your payoff will not be reduced, you will receive the full payoff, i.e. your Reported Number x \euro{15}.
    If the experimenter [the algorithm] decides that your payoff will be reduced,  you will receive a reduced payout depending on the number on your card - Number on Card x \euro{7.50}. Accordingly, possible winning amounts would be:
        \begin{table}[H]
            \centering
            \begin{tabular}{|c|c|}
                \hline
                \textbf{Number on card} & \textbf{Additional prize} \\
                \hline
                1 & \euro{7.50} \\
                2 & \euro{15.00} \\
                3 & \euro{22.50} \\
                4 & \euro{30.00} \\
                5 & \euro{37.50} \\
                6 & \euro{45.00} \\
                \hline
            \end{tabular}
            \label{tab:reduced_payoffs_bb}
        \end{table}
    \item This means, you cannot go away empty-handed if you are drawn for the additional prize.
    \item In both cases the experiment is finished afterwards.
    \item To summarize: The experimenter [algorithm] makes at least 1 and max. 2 decisions during the payout process.
\end{itemize}

\section*{Additional remarks}

\begin{itemize}
    \item No communication is allowed during the experiment.
    \item All decisions you make during this experiment will be completely anonymous. None of the other participants will learn of your identity, the decisions you make, or the payoff you receive. The data will be analyzed for scientific purposes only.
\end{itemize}

\textbf{Good luck and thank you for participating in this experiment!}

\clearpage

\section*{Comprehension questions}

\subsection*{Please answer the following questions.}
\textit{[Questions asked in all treatments, translated form German, 'X' indicates correct answer]}

Who will receive a bonus payment?
\begin{table}[H]
    \begin{tabular}{c l}
         $\ocircle$ & All participants will receive a bonus payment. \\
         $\ocircle$ & One half of the participants will receive a bonus payment. \\
         $\ocircle$ & One participant will receive a bonus payment. X \\
    \end{tabular}
    \label{tab:comp1}
\end{table}

Which statement regarding the payoff process is correct?
\begin{table}[H]
    \begin{tabular}{c l}
         $\ocircle$ & The payout of the bonus payment takes place in camera. X \\
         $\ocircle$ & The amount of the bonus payment exclusively depends on the reported number. \\
         $\ocircle$ & The amount of the bonus payment is fixed. \\
    \end{tabular}
    \label{tab:comp2}
\end{table}

Which payment amount total is the minimum you will receive in case you are drawn to receive the bonus payment?
\begin{table}[H]
    \begin{tabular}{c l}
         $\ocircle$ & \euro{7.50} (i.e., the fixed payment) \\
         $\ocircle$ & \euro{15.00} (i.e., the fixed payment + \euro{7.50}) X\\
         $\ocircle$ & \euro{22.50} (i.e., the fixed payment + \euro{15}) \\
    \end{tabular}
    \label{tab:comp3}
\end{table}

What happens in case your card is inspected?
\begin{table}[H]
    \begin{tabular}{c l}
         $\ocircle$ & You receive a new card. \\
         $\ocircle$ & You only receive the show-up fee. \\
         $\ocircle$ & Your reported number will be compared with the number on your card. X \\
    \end{tabular}
    \label{tab:comp4}
\end{table}

\subsection*{Please answer the following questions.}
\textit{[Questions asked in human treatments, translated form German, 'X' indicates correct answer]}

Which statement about Lottery 1 is correct?
\begin{table}[H]
    \begin{tabular}{c l}
         $\ocircle$ & All participants play Lottery 1. \\
         $\ocircle$ & The drawable numbers from 1 to 10 have different probabilities. \\
         $\ocircle$ & Since a card check takes place if the number drawn in Lottery 1 is lower than or equal to the number  \\
          & you reported, the higher your reported number, the higher the probability of your card getting checked. X
    \end{tabular}
    \label{tab:comp5}
\end{table}

Which statement regarding the card check is correct?
\begin{table}[H]
    \begin{tabular}{c l}
         $\ocircle$ & Whether a check takes place is decided by yourself. \\
         $\ocircle$ & If your reported number does not match the number on your card in a check, Lottery 2 \\
          & follows. You still have a chance of receiving the full payoff (i.e., your reported number x \euro{15}). X\\
         $\ocircle$ & If your reported number does not match the number on your card in a check, you receive a \\
          & reduced payoff (i.e., the number you drew x \euro{7.50}). \\
    \end{tabular}
    \label{tab:comp6}
\end{table}

Which statement about Lottery 2 is correct?
\begin{table}[H]
    \begin{tabular}{c l}
         $\ocircle$ & If the number drawn in Lottery 2 (between 1 and the number you reported) is higher than the \\
          & number on your card, your payoff is reduced (to 'number on your card' x \euro{7.50}).  X \\
         $\ocircle$ & Every prize draw winner plays Lottery 2. \\
         $\ocircle$ & If the number drawn in Lottery 2 (between 1 and the number you reported) is lower than or \\
          & equal to the number on your card, your payoff is reduced (to 'number on your card' x \euro{7.50}). \\
    \end{tabular}
    \label{tab:comp7}
\end{table}

\subsection*{Please answer the following questions.}
\textit{[Questions asked in machine treatments, translated form German, 'X' indicates correct answer)]}

Which of the following statements regarding Decision 1 \& 2 is correct?
\begin{table}[H]
    \begin{tabular}{c l}
         $\ocircle$ & Decision 1 is made for all participants. \\
         $\ocircle$ & Decision 2 is always made for the winner. \\
         $\ocircle$ & Decision 1 decides whether a card check will take place. X \\
    \end{tabular}
    \label{tab:comp5bb}
\end{table}

Which of the following statements regarding the card check is correct?
\begin{table}[H]
    \begin{tabular}{c l}
         $\ocircle$ & Even if the number on your card does not match the number you reported, you can still receive \\
          & the full payoff (i.e., your reported number x \euro{15}). X \\
         $\ocircle$ & Even if the number on your card matches the number you reported, Decision 2 follows. \\
         $\ocircle$ & If your reported number matches the number on your card, you will receive a payoff in the \\ 
          & amount of your drawn number x \euro{7.50}. \\
    \end{tabular}
    \label{tab:comp6bb}
\end{table}

%% file: hmd_questionnaire.tex
\section{Questionnaire} \label{hmd_questionnaire}

Thank you for putting in your number. You will find out whether you receive the additional prize at the end of the experiment.

In the following, we ask you to fill out our multi-part questionnaire. There is no "right" or "wrong" here. Simply answer the questions in the way that seems most appropriate to you personally. The questionnaire consists of 7 parts in total, each containing a different number of questions.

Your answers will be treated completely anonymously and will not affect your chances of winning.

\subsection*{Please answer the following questions.}

Please recall again the verification process described for the previous decision.

Which of the following entities would you prefer to be audited by in this process?

\begin{table}[H]
    \begin{tabular}{c l}
         $\ocircle$ & a human \\
         $\ocircle$ & a machine (e.g. algorithm, AI, computer program, ...) \\
    \end{tabular}
    \label{tab:hmd_preference}
\end{table}

Which of the following entities do you consider to have more decision discretion? 

\begin{table}[H]
    \begin{tabular}{c l}
         $\ocircle$ & a human \\
         $\ocircle$ & a machine (e.g. algorithm, AI, computer program, ...) \\
    \end{tabular}
    \label{tab:hmd_discretionexp}
\end{table}

Which of the following entities do you consider more prone to making mistakes/errors?

\begin{table}[H]
    \begin{tabular}{c l}
         $\ocircle$ & a human \\
         $\ocircle$ & a machine (e.g. algorithm, AI, computer program, ...) \\
    \end{tabular}
    \label{tab:hmd_errorexp}
\end{table}

\textbf{Please note: Your answers have no effect on your chance of winning the prize draw.}

\subsection*{Please answer the following questions.}

In the following questionnaire, we will ask you about your interaction with technical systems. The term “technical systems” refers to apps and other software applications, as well as entire digital devices (e.g., mobile phone, computer, TV, car navigation).
\par
Please indicate to what extent you agree to the following statements.

\textbf{Please note: Your answers have no effect on your chance of winning the prize draw.}

\begin{table}[H]
    \centering
    \begin{adjustbox}{max width=\textwidth}
    \begin{tabular}{l c c c c c c}
    \hline
        & \makecell{Completely \\ disagree} & \makecell{Largely \\ disagree}  & \makecell{Slightly\\ disagree } & \makecell{Slightly \\agree} & \makecell{Largely \\ agree} & \makecell{Completely \\ agree } \\
    \hline
         I like to occupy myself in greater detail with technical systems. & $\ocircle$ & $\ocircle$ & $\ocircle$ & $\ocircle$ & $\ocircle$ & $\ocircle$ \\
         I like testing the functions of new technical systems. & $\ocircle$ & $\ocircle$ & $\ocircle$ & $\ocircle$ & $\ocircle$ & $\ocircle$ \\
         I predominantly deal with technical systems because I have to. & $\ocircle$ & $\ocircle$ & $\ocircle$ & $\ocircle$ & $\ocircle$ & $\ocircle$ \\
         When I have a new technical system in front of me, I try it out intensively. & $\ocircle$ & $\ocircle$ & $\ocircle$ & $\ocircle$ & $\ocircle$ & $\ocircle$ \\
         I enjoy spending time becoming acquainted with a new technical system. & $\ocircle$ & $\ocircle$ & $\ocircle$ & $\ocircle$ & $\ocircle$ & $\ocircle$ \\
         It is enough for me that a technical system works; I don’t care how or why. & $\ocircle$ & $\ocircle$ & $\ocircle$ & $\ocircle$ & $\ocircle$ & $\ocircle$ \\
         I try to understand how a technical system exactly works. & $\ocircle$ & $\ocircle$ & $\ocircle$ & $\ocircle$ & $\ocircle$ & $\ocircle$ \\
         It is enough for me to know the basic functions of a technical system. & $\ocircle$ & $\ocircle$ & $\ocircle$ & $\ocircle$ & $\ocircle$ & $\ocircle$ \\
         I try to make full use of the capabilities of a technical system. & $\ocircle$ & $\ocircle$ & $\ocircle$ & $\ocircle$ & $\ocircle$ & $\ocircle$ \\         
    \hline
    \end{tabular}
    \end{adjustbox}
    \label{tab:hmd_ati}
\end{table}

\subsection*{Please answer the following questions.}

For the following statements, please indicate to what extent you consider the actions or behaviours described to be ethically problematic.
Please indicate your assessment between "Definitely not problematic" and "Definitely problematic" on the following scale:

\textbf{Please note: Your answers have no effect on your chance of winning the prize draw.}

\begin{table}[H]
    \centering
    \begin{adjustbox}{max width=\textwidth}
    \begin{tabular}{l c c c c c}
    \hline
        & \makecell{Definitely \\ unproblematic} & \makecell{Rather \\ unproblematic}  & \makecell{Not \\ sure} & \makecell{Rather \\problematic} & \makecell{Definitely \\ problematic} \\
    \hline
         Taking some questionable deductions on your income tax return. & $\ocircle$ & $\ocircle$ & $\ocircle$ & $\ocircle$ & $\ocircle$\\
         Having an affair with a married man/woman. & $\ocircle$ & $\ocircle$ & $\ocircle$ & $\ocircle$ & $\ocircle$\\
         Passing off somebody else’s work as your own. & $\ocircle$ & $\ocircle$ & $\ocircle$ & $\ocircle$ & $\ocircle$\\
         Revealing a friend’s secret to someone else.  & $\ocircle$ & $\ocircle$ & $\ocircle$ & $\ocircle$ & $\ocircle$\\
         Leaving your young children alone at home while running an errand. & $\ocircle$ & $\ocircle$ & $\ocircle$ & $\ocircle$ & $\ocircle$\\
         Not returning a wallet you found that contains 200€. & $\ocircle$ & $\ocircle$ & $\ocircle$ & $\ocircle$ & $\ocircle$\\ 
    \hline
    \end{tabular}
    \end{adjustbox}
    \label{tab:hmd_dospert}
\end{table}

\subsection*{Please answer the following questions.} 

What is your age? \\

What is your gender?
\begin{table}[H]
    \begin{tabular}{c l}
         $\ocircle$ & Male \\
         $\ocircle$ & Female \\
         $\ocircle$ & Non-binary \\
    \end{tabular}
    \label{tab:hmd_gender}
\end{table}

What is your current study major? \\

In general, how willing are you to take risks?

Not at all willing to take risks $\ocircle$ $\ocircle$ $\ocircle$ $\ocircle$ $\ocircle$ $\ocircle$ $\ocircle$ $\ocircle$ $\ocircle$ $\ocircle$ $\ocircle$ Very willing to take risks \\

Is there anything else you would like to tell us? (optional)

%% file: hmd_derivation.tex
\section{Derivation of payoff utility function} \label{hmd_derivation}

\begin{table}[H]
    \centering
    \caption{Derivation of payoff-maximizing decision strategy}
    \label{tab:hmd_derivation}
    \begin{threeparttable}
    \begin{tabular}{c l c c c c c c }
         \hline
         \hline
         & & \multicolumn{6}{c}{\textbf{Report}}\\ \cline{3-8}
         & & 1 & 2 & 3 & 4 & 5 & 6\\
         \hline
         \textbf{Number drawn} & & & & & & & \\
         \hline
         \multirow{7}{*}{\textbf{1}} & P(audit) & 0.1 & 0.2 & 0.3 & 0.4 & 0.5 & 0.6 \\
         & P(punishment$|$audit) & 0 & 0.5 & 0.667 & 0.75 & 0.8 & 0.833\\
         & P(punishment) & 0 & 0.1 & 0.2 & 0.3 & 0.4 & 0.5\\
         & P(cheating successful) & 0 & 0.9 & 0.8 & 0.7 & 0.6 & 0.5\\
         & E(payoff) & 15 & 27.75 & 37.5 & 44.25 & 48 & \textbf{48.75}\\
         & U(cheating) &  & 12.75 & 22.5 & 29.25 & 33 & 33.75\\
         & U'(cheating) & & 12.75 & 9.75 & 6.75 & 3.75 & 0.75\\
         \hline
         \multirow{7}{*}{\textbf{2}} & P(audit) & & 0.2 & 0.3 & 0.4 & 0.5 & 0.6 \\
         & P(punishment$|$audit) & & 0 & 0.333 & 0.5 & 0.6 & 0.667\\
         & P(punishment) & & 0 & 0.1 & 0.2 & 0.3 & 0.4 \\
         & P(cheating successful) & & & 0.9 & 0.8 & 0.7 & 0.6\\
         & E(payoff) & & & 42.00 & 51.00 & 57.00 & \textbf{60.00}\\
         & U(cheating) & & & 12 & 21 & 27 & 30\\
         & U'(cheating) & & & 12 & 9 & 6 & 3\\
         \hline
         \multirow{7}{*}{\textbf{3}} & P(audit) & & & 0.3 & 0.4 & 0.5 & 0.6 \\
         & P(punishment$|$audit) & & & & 0.25 & 0.4 & 0.5\\
         & P(punishment) & & & & 0.1 & 0.2 & 0.3 \\
         & P(cheating successful) & & & & 0.9 & 0.8 & 0.7\\
         & E(payoff) & & & & 56.25 & 64.50 & \textbf{69.75}\\
         & U(cheating) & & & & 11.25 & 19.5 & 24.75\\
         & U'(cheating) & & & & 11.25 & 8.25 & 5.25\\
         \hline
         \multirow{7}{*}{\textbf{4}} & P(audit) & & & & 0.4 & 0.5 & 0.6 \\
         & P(punishment$|$audit) & & & & 0 & 0.2 & 0.333 \\
         & P(punishment) & & & & 0 & 0.1 & 0.2\\
         & P(cheating successful) & & & & 0 & 0.9 & 0.8\\
         & E(payoff) & & & & 60 & 70.50 & \textbf{78.00}\\
         & U(cheating) & & & & & 10.50 & 18\\
         & U'(cheating) & & & & & 10.50 & 7.50\\
         \hline
         \multirow{7}{*}{\textbf{5}} & P(audit) & & & & & 0.5 & 0.6 \\
         & P(punishment$|$audit) & & & & & 0 & 0.167\\
         & P(punishment) & & & & & 0 & 0.1\\
         & P(cheating successful) & & & & & 0 & 0.9 \\
         & E(payoff) & & & & & 75 & \textbf{84.75} \\
         & U(cheating) & & & & &  & 9.75\\
         & U'(cheating) & & & & &  & 9.75\\
         \hline
         \multirow{7}{*}{\textbf{6}} & P(audit) & & & & & & 0.6 \\
         & P(punishment$|$audit) & & & & & & 0 \\
         & P(punishment) & & & & & & 0\\
         & P(cheating successful) & & & & & & 0\\
         & E(payoff) & & & & & & 90\\
         & U(cheating) & & & & & & \\
         & U'(cheating) & & & & & & \\
         \hline
         \hline
    \end{tabular}
    \begin{tablenotes}
        \small \item \textit{Note:} Abbreviations in table denote the following: $P(x)$: probability; $P(x_1 | x_2)$:  conditional probability; $E(x)$: expected value; $U(x)$: utility; $U'(x)$: marginal utility.
    \end{tablenotes}
    \end{threeparttable}
\end{table}

%% file: hmd_audit.tex
\section{Verification process illustrations and outcomes} \label{hmd_audit}

\subsubsection*{Human verification}

\begin{figure}[H]
  \includegraphics[scale=0.35]{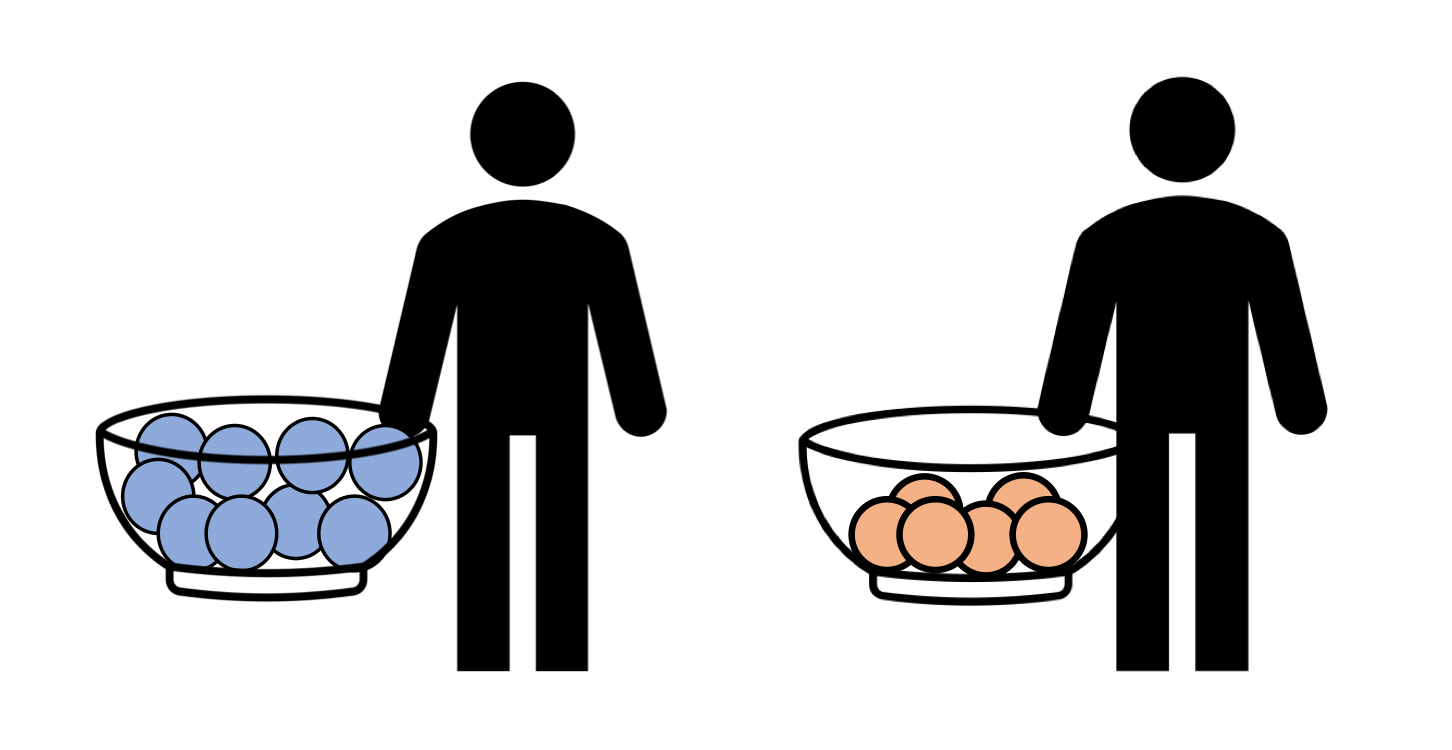}
  \centering
  \caption{Illustration of human verification process}
  \label{fig:hmd_human_audit}
\end{figure}

The pictograms displayed in Figure \ref{fig:hmd_human_audit} were included in the experimental instructions to illustrate the Verification Part conducted by a human.

\vspace{1cm}

\subsubsection*{Algorithmic verification}

\begin{figure}[H]
  \includegraphics[scale=0.3]{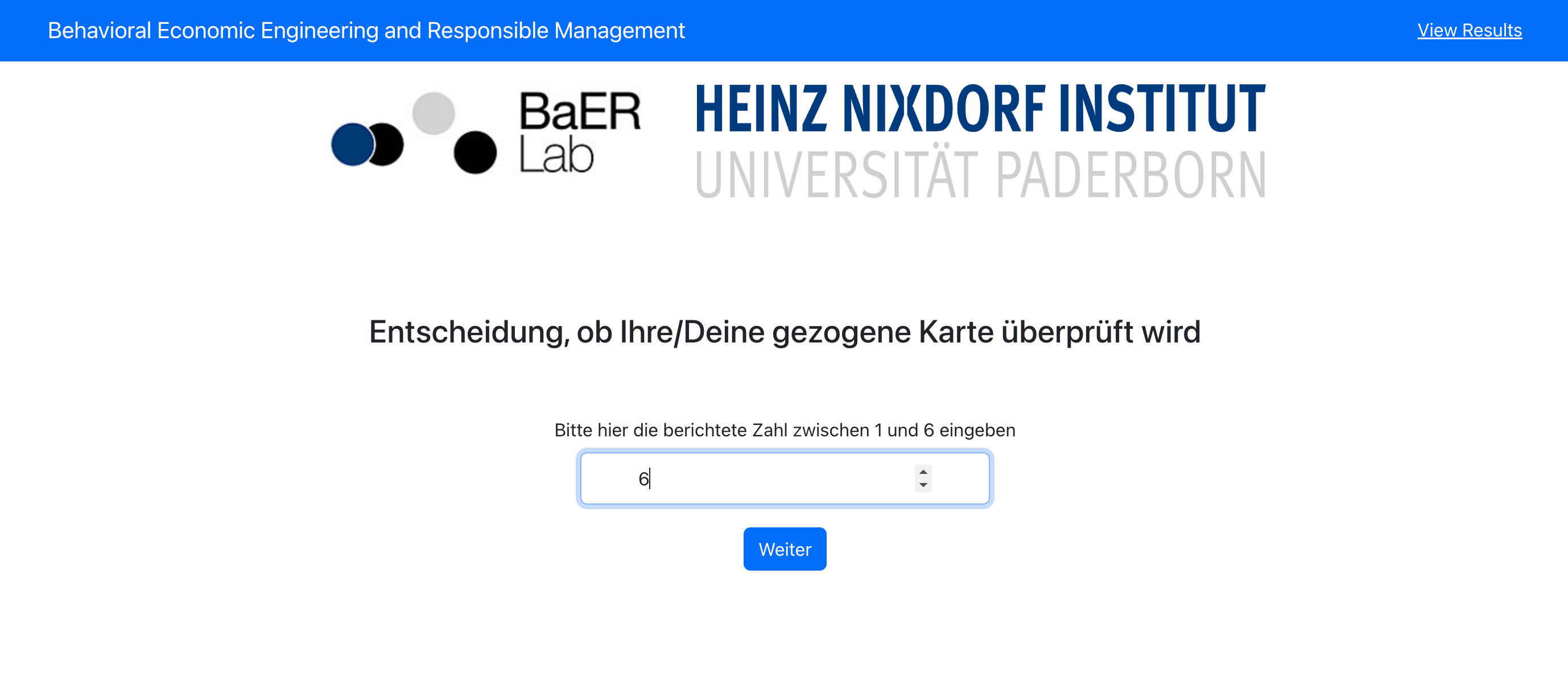}
  \centering
  \caption{Example for the algorithmic verification interface: Input number 6 (notation in German)}
  \label{fig:hmd_algo_audit_example_input}
\end{figure}

Translation of text in Figure \ref{fig:hmd_algo_audit_example_input}: 

\textit{"Decision whether your drawn card is checked"}

\textit{"Please enter your reported number:"}

\clearpage

\begin{figure}[H]
  \includegraphics[scale=0.3]{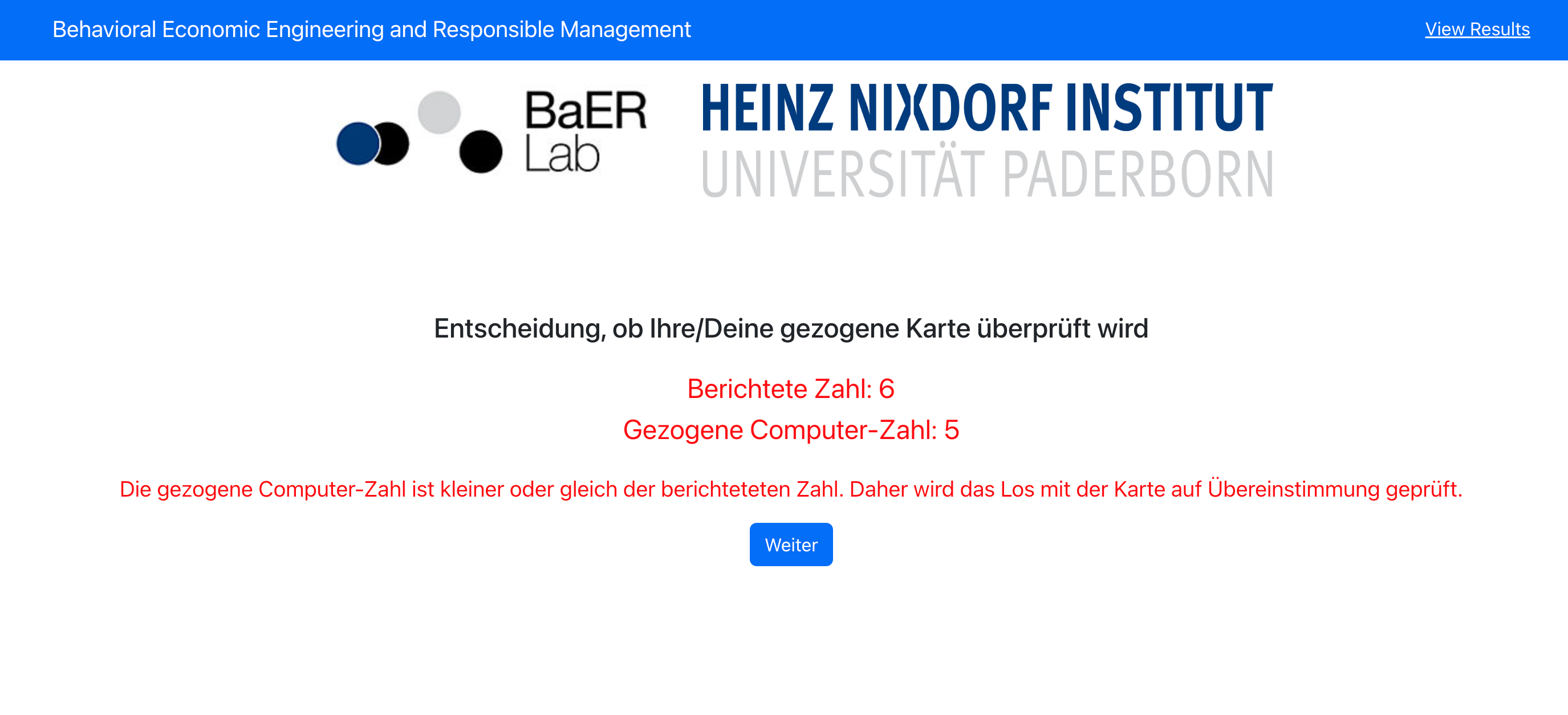}
  \centering
  \caption{\small Example for the algorithmic verification interface: Decision on inspection of drawn card (notation in German)}
  \label{fig:hmd_algo_audit_example_decision}
\end{figure}

Translation of text in Figure \ref{fig:hmd_algo_audit_example_decision}: 

\textit{"Decision whether your drawn card is checked"}

\textit{"Reported number: 6:"}

\textit{"Number drawn by the computer: 5"}

\textit{"The computer's number is lower than or equal to your reported number. Therefore, your drawn card will be checked."}

\clearpage

\begin{figure}[H]
  \includegraphics[scale=0.3]{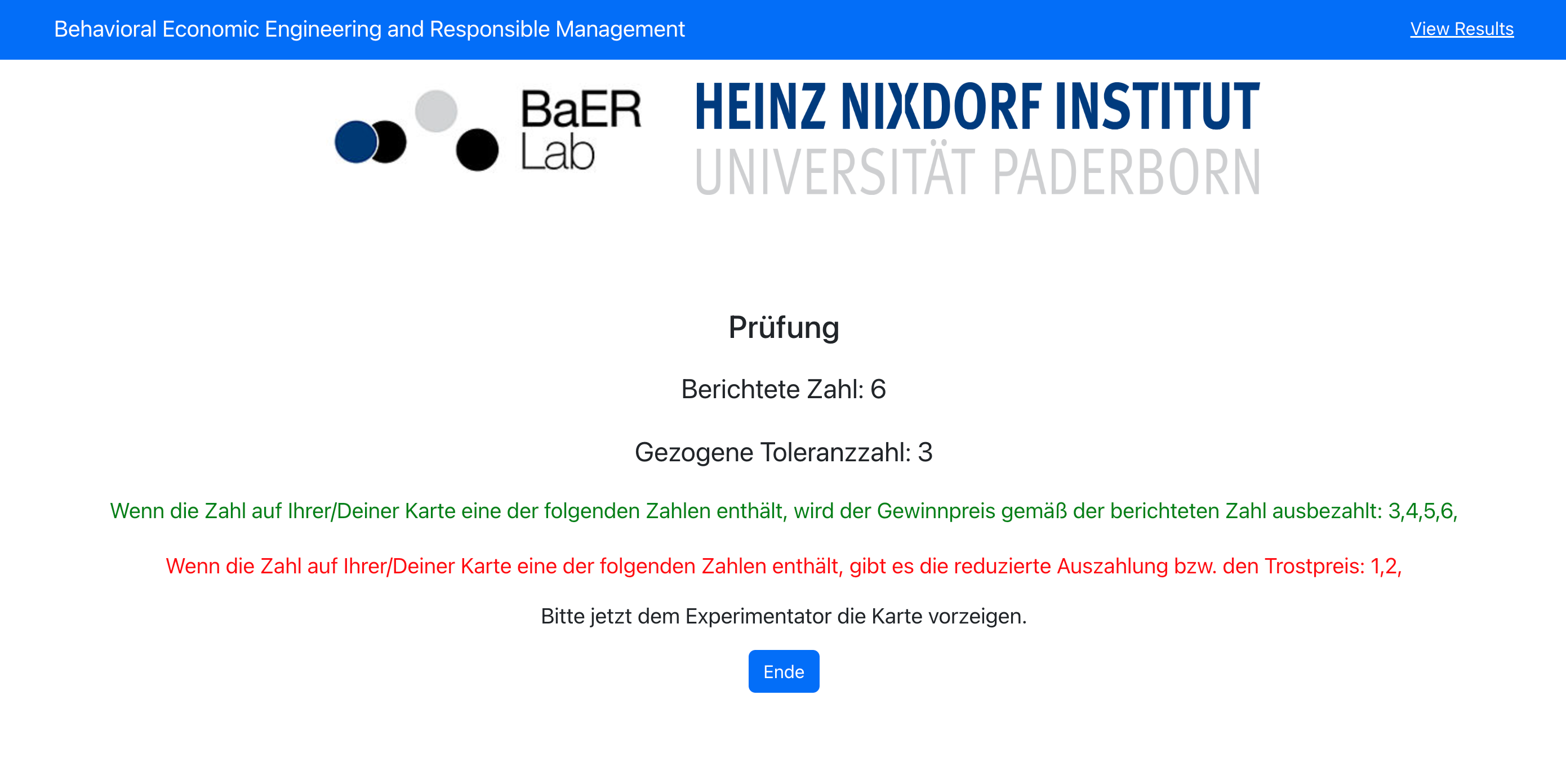}
  \centering
  \caption{\small Example for the algorithmic verification interface: Decision on payoff reduction (notation in German)}
  \label{fig:hmd_algo_audit_example_tolerance}
\end{figure}

Translation of text in Figure \ref{fig:hmd_algo_audit_example_tolerance}: 

\textit{"Check"}

\textit{"Reported number: 6"}

\textit{"Tolerance number: 3"}

\textit{"If the number on your card is one of the following, you receive the payoff according to your reported number: 3, 4, 5, 6"}

\textit{"If the number on your card is one of the following, your payoff is reduced: 1, 2"}

\textit{"Please show your card to the experimenter now."}

\vspace{5mm}

\subsubsection*{Audit outcomes by session}

Table \ref{tab:hmd_session_audits} displays event sequences and outcomes of the Verification Part for each experiment session.

\begin{table}[H]
    \centering
    \caption{Verification process, by session}
    \begin{tabular}{ c c c c c c c }
         \hline
         \hline
         \textbf{Session} & \textbf{Reported} & \textbf{Lottery 1} & \textbf{Card checked} & \textbf{Lottery 2} & \textbf{Reduction} & \textbf{Final payoff} \\
         \hline
         1 & 2 & 3 & No & - & - & \euro{30}\\
         2 & 4 & 6 & No & - & - & \euro{60}\\
         3 & 2 & 7 & No & - & - & \euro{30}\\
         4 & 6 & 4 & Yes & 1 & No & \euro{90} \\
         5 & 2 & 5 & No & - & - & \euro{30}\\
         6 & 2 & 1 & No & - & - & \euro{30} \\
         7 & 5 & 7 & Yes & 4 & Yes & \euro{15}\\
         8 & 2 & 5 & No & - & - & \euro{30} \\
         9 & 6 & 4 & Yes & 4 & Yes & \euro{15}\\
         10 & 2 & 8 & No & - & - & \euro{30} \\         
         \hline
         \hline
    \end{tabular}
    \label{tab:hmd_session_audits}
\end{table}